\documentclass[pre,onecolumn,superscriptaddress,amsmath,amssymb]{revtex4-1}
\usepackage{tikz}
\usepackage[]{marvosym}
\usepackage{enumitem}
\usepackage[toc,page]{appendix}
\usepackage{graphicx}
\usepackage{dcolumn}
\usepackage{bm}
\usepackage{natbib}
\usepackage{float}
\usepackage{hyperref}
\usepackage{url}

\newcommand{\modif}[1]{\textcolor{black}{#1}} 

\begin{document}
\preprint{APS/123-QED}
\bibliographystyle{apsrev4-1}
\title{Fast determination of \modif{coarse grained }cell anisotropy and size\\
in epithelial tissue images using Fourier transform}

\author{M. Durande (\Letter)}%
 \affiliation{Laboratoire Mati\`ere et Syst\`emes Complexes, Universit\'e Denis Diderot - Paris 7, CNRS UMR 7057, 10 rue Alice Domon et L\'eonie Duquet, F-75205 Paris Cedex 13, France}
\affiliation{Univ. Lyon, Universit\'e Claude Bernard Lyon 1, CNRS UMR 5306,  Institut Lumi\`ere Mati\`ere, Campus LyonTech - La Doua,  Kastler building, 10 rue Ada Byron,  
F-69622 Villeurbanne Cedex, France}%

\author{S. Tlili}
 \affiliation{Laboratoire Mati\`ere et Syst\`emes Complexes, Universit\'e Denis Diderot - Paris 7, CNRS UMR 7057, 10 rue Alice Domon et L\'eonie Duquet, F-75205 Paris Cedex 13, France}
 \affiliation{Mechanobiology Institute, Department of Biological Sciences, National University of Singapore, 5A Engineering Drive, 1, 117411 Singapore}
 
\author{T. Homan}%
\affiliation{Univ. Lyon, Universit\'e Claude Bernard Lyon 1, CNRS UMR 5306,  Institut Lumi\`ere Mati\`ere, Campus LyonTech - La Doua,  Kastler building, 10 rue Ada Byron,  
F-69622 Villeurbanne Cedex, France}%

\author{B. Guirao}%
\affiliation{Polarity, Division and Morphogenesis Team, Institut Curie, CNRS UMR 3215, INSERM U934, 26 rue d'Ulm, 75248 Paris Cedex 05, France}%

\author{F. Graner (\Letter)}%
\affiliation{Laboratoire Mati\`ere et Syst\`emes Complexes, Universit\'e Denis Diderot - Paris 7, CNRS UMR 7057, 10 rue Alice Domon et L\'eonie Duquet, F-75205 Paris Cedex 13, France}%

\author{H. Delano\"e-Ayari (\Letter)}%
\affiliation{Univ. Lyon, Universit\'e Claude Bernard Lyon 1, CNRS UMR 5306,  Institut Lumi\`ere Mati\`ere, Campus LyonTech - La Doua,  Kastler building, 10 rue Ada Byron,  
F-69622 Villeurbanne Cedex, France}%


\begin{abstract}

Mechanical strain and stress play a major role in biological processes such as wound healing or morphogenesis. 
To assess this role quantitatively, fixed or live images of tissues  are acquired  at a cellular precision in large fields of views. To exploit these data, large \modif{numbers} of cells have to be analyzed to extract  cell shape anisotropy and cell size. Most frequently, this is performed through detailed individual cell contour determination, using  so-called segmentation computer programs, complemented if necessary by manual detection and error corrections.  
 However, a \modif{coarse grained} and faster technique can be recommended in at least three situations. First, when detailed information on individual cell contours is not required, for instance in studies which require only coarse-grained average information on cell anisotropy. Second, as an exploratory step to determine whether full segmentation can be potentially useful. Third, when segmentation is too difficult, for instance due to poor image quality or too large \modif{a} cell number. We developed a user-friendly, Fourier \modif{transform-based} image analysis pipeline. It is fast (typically $10^4$ cells per minute with a current laptop computer) and suitable for time, space or ensemble averages. We validate it on one set of artificial images and on two sets of fully segmented images, \modif{one} from \modif{a} Drosophila pupa and \modif{the other from} \modif{a} chicken embryo; the pipeline results are robust. Perspectives include \textit{in vitro} tissues, non-biological cellular patterns such as foams, and \modif{$xyz$ stacks}.

\end{abstract}

\maketitle


\section{Introduction}
\label{sec:Introduction}

During important physiological processes such as wound healing, morphogenesis or metastasis, 
cells  deform, migrate, exchange neighbors, divide and die. A proper mechanical description of such complex active system requires the characterization of cell size, cell shape and changes thereof \cite{Heisenberg2013}.
Fluorescent labeling of cell contours and \modif{progress} in microscopy have led to the acquisition of large  tissue images with high signal-to-noise ratio.
Determination of individual cell contours have allowed \modif{the application of} mechanical approaches based on quantitative data analysis of cell packing within epithelial tissues \cite{Zallen2004,Classen2005,Hayashi2004}; \modif{the development of} quantitative modeling of  tissue structure \cite{Kafer2007,Hilgenfeldt2008,Farhadifar2007}; and even \modif{the linking of} cell-level changes to morphogenetic movements \cite{Rauzi2008,Blanchard2009,Guirao2015,Etournay2016}. While these  studies were bidimensional, three dimensional studies are becoming increasingly common \cite{Faure2016,Sherrard2010,Diaz-de-la-Loza2018}.

\begin{figure}[h]
\includegraphics[scale=0.8]{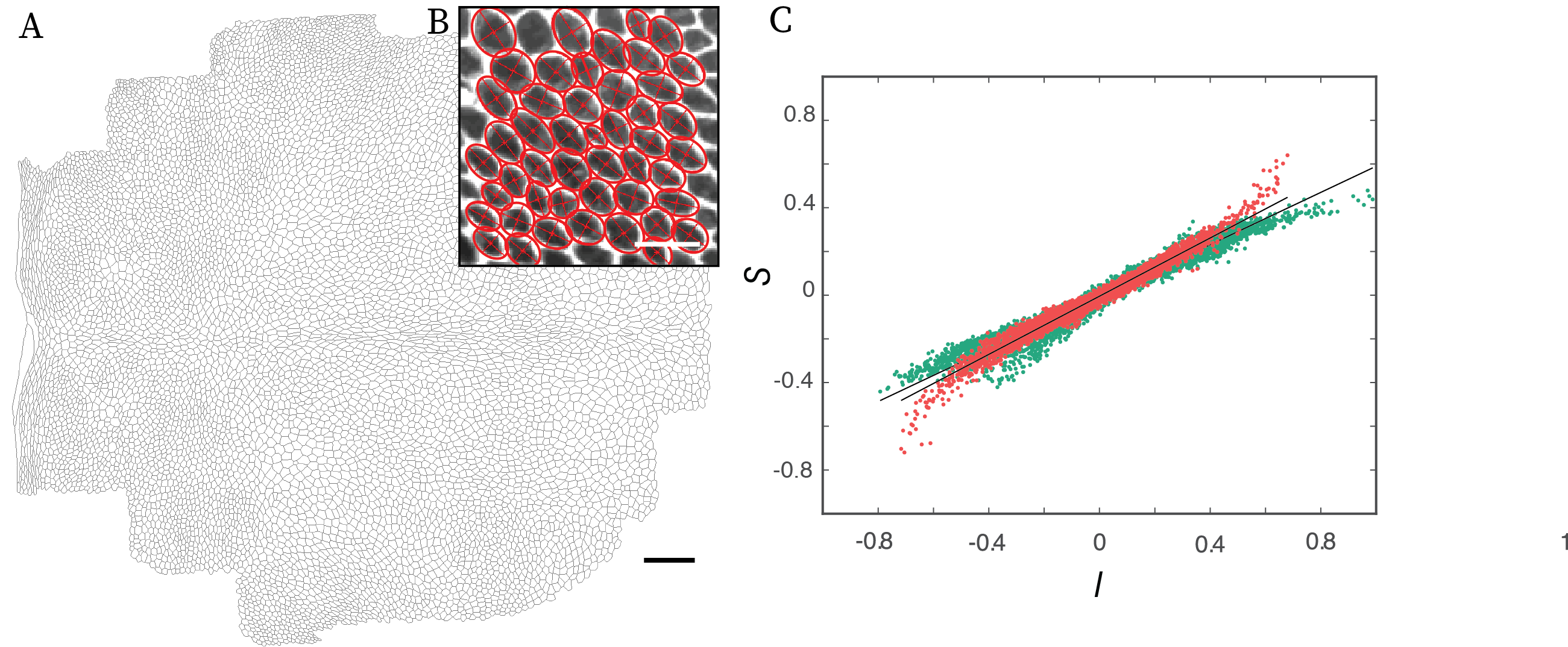}
\caption{ 
Cell segmentation \modif{reveals} a correlation between cell shape and elastic stress anisotropies. 
$(A)$ Whole \modif{segmented image} of the Drosophila dorsal thorax during its metamorphosis. Head is on the right of the image and abdomen on the left. Scale bar is 50 $\mu$m. $(B)$ Sub-image in original grey levels, scale bar is 10 $\mu$m. 
Cells are transfected to label  the membrane  with a fluorescent marker (green fluorescent protein, GFP),  imaged using an inverted confocal spinning disk microscope, and segmented using a home-made software followed by automatic and manual rounds of correction  \cite{Guirao2015}. 
In $B$, the inertia matrix of each cell is superimposed as an ellipse. (C) Diagonal component (green) and off-diagonal component (red) of the cell-cell junction tension contribution to stress $S$, representing elastic stress anisotropy, versus the corresponding anisotropic (diagonal and off-diagonal) component of the inertia matrix $I$, representing cell shape anisotropy. Correlation coefficients are respectively $0.97$ and $0.96$, slopes are respectively $0.6,0.67$. Tensors built with data from  \cite{Guirao2015}, here plotted after adimensionalisation by the isotropic part of the respective tensor. To compute each of the 14112 points, tensors of individual cells are computed before being averaged in Eulerian grids of $40 \times 40$~$\mu$m$^2$ with 50\% overlap. Then a sliding average is performed on 2 h (24 frames) \modif{time windows} with a one hour overlap. Boxes at the pattern boundary which are filled at less than 30\% by cells are excluded from the fit.}
\label{fig:pup}
\end{figure}

These studies have been performed successfully using detailed determination of cell contours, also called ``cell segmentation",  either manual, automatic or a combination of both. Segmentation programs are available in an increasing number (see \cite{Faure2016,Heller2016,Etournay2016} and references therein). 
Fig.~\ref{fig:pup} presents an example of such segmentation, in which the dorsal thorax  of a fruit fly ({\it Drosophila}) is imaged at the \modif{pupal stage}, i.e. during its metamorphosis from larva to adult. 
For each cell, the inertia matrix (see Appendix \ref{sec:def}) is computed and represented as an ellipse which fits the cell contour. 
It is completely defined by three parameters: its major and minor axis length ($L_{maj}$, $L_{min}$) and the orientation $\theta$ of its major axis with respect to the $x$-axis.
The inertia matrix can be averaged on any given region, yielding  an average inertia matrix and thus an ellipse characterising the average cell size and shape in this region.
The cell area is characterised by $\pi L_{maj} L_{min}$.

The ratio $L_{maj}/L_{min}$ and the angle $\theta$ characterize the anisotropic part of the cell shape and are expressed by the anisotropic part of the inertia matrix. Interestingly, it is shown with image analysis using force inference on detailed cell contours \cite{Guirao2015} that the anisotropic part of the inertia matrix correlates strongly with the anisotropic part of the stress at cell-cell junctions (Fig.~\ref{fig:pup}C).
This last result is in agreement with a recent theoretical prediction 
\cite{Ishihara2017a} (under the assumption that cell-cell junctions and sizes are homogeneous in the tissue)
and suggests \modif{that shape measurements could be used as a proxy to estimate stress (with exceptions recently documented in some extreme cases, see \cite{Latorre2018})}. Moreover, the inertia matrix also correlates strongly with the texture tensor (See Appendix A, Fig.~\ref{fig:MI}) that is used to \modif{statistically} define the  strain \cite{Graner2008}.
This reinforces the interest of cell shape measurements, as an approximative but fast and simple alternative to stress measurements. Since stress is defined as a coarse grained quantity over a tissue region, average shape measurements should suffice without need for detailed individual cell shape segmentation. 

There are cases where a segmentation-free method of cell shape determination is potentially useful. For instance, a fast exploration of cell shape variation in time and space could help determine its role in a given biological question, before undertaking the detailed segmentation. Or, it could  partially replace segmentation in cases where the image quality makes it difficult to segment with reasonable effort and sufficient precision: 
low or variable contrast, low signal to noise ratio, interrupted cell edges, large variability of cell sizes, variety of cell types or very contorted cell shapes. 
Even when the image can be segmented, the cell number can be \modif{much} too large to enable segmentation within a reasonable amount of time.

\begin{figure}[h]
\includegraphics[scale = 0.8]{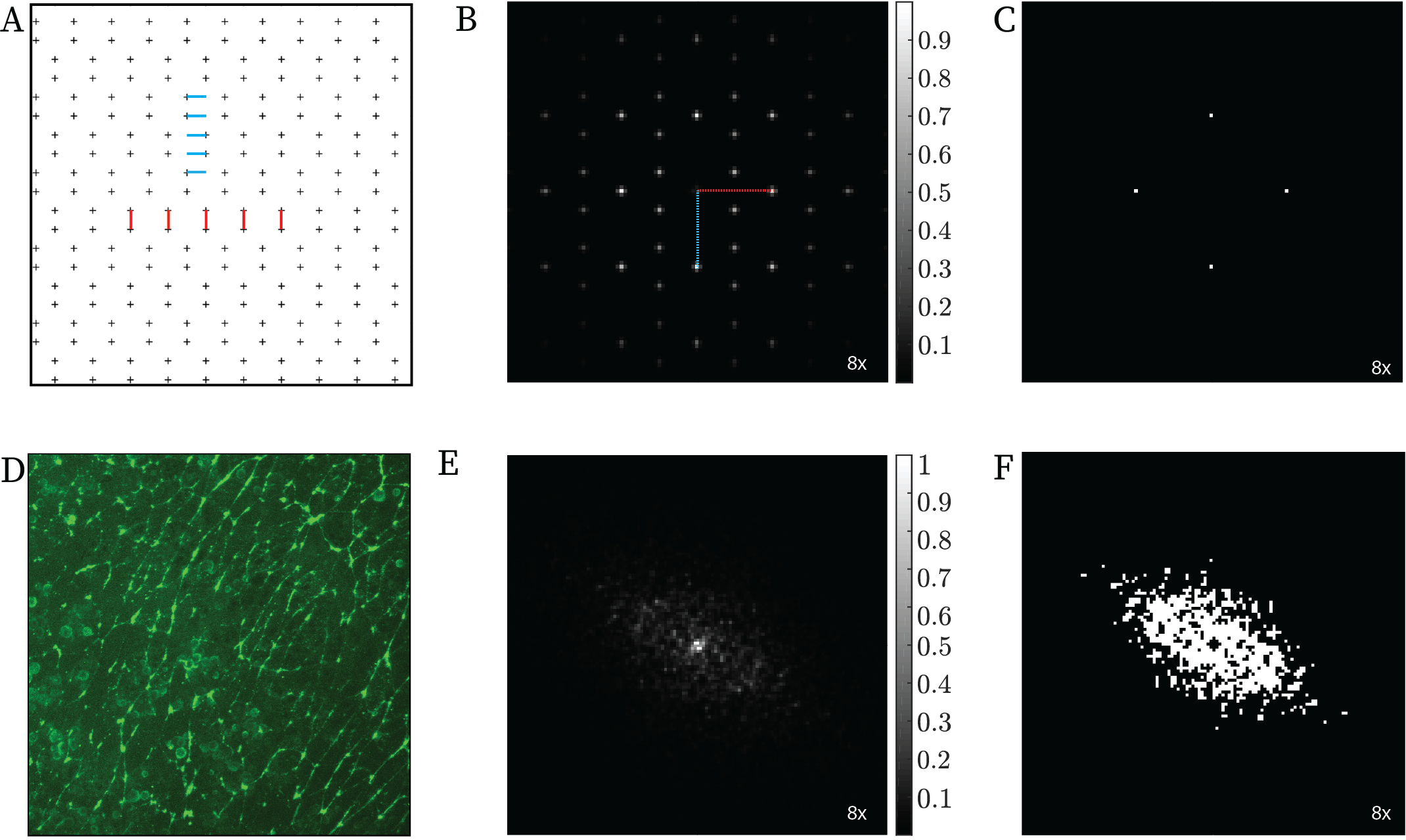}
\caption{
Fourier transform examples. $(A)$ Points \modif{arranged} \modif{in a spatially periodic array. Blue and red bars highlight some periods.} $(B)$ Fourier spectrum of $A$ with Gaussian blur of \modif{standard deviation 0.6}.
$(C)$ Same as $B$ after selecting \modif{a proportion $p$ of the image pixels which are the brightest, with $p=7\cdot 10^{-6}$}.
The two principal directions are visible. $(D)$ Anisotropic myosin distribution in \modif{a} chicken embryo during morphogenesis, courtesy of \modif{C.J.}  Weijer. $(E)$ Fourier spectrum of $D$ with Gaussian blur of \modif{standard deviation 0.6}.
$(F)$ Same as $E$ after selecting \modif{a proportion $p$ of the image pixels which are the brightest, with $p=10^{-3}$}.
\modif{In $B,C,E,F$,}
``8x" signifies that the \modif{spectra} are zoomed 8 times.}
\label{im:diffTF}
\end{figure}

 Different techniques have been probed to quantify a pattern anisotropy without segmentation, such as Hough transform  \cite{Duda1972}, Radon transform \cite{Streichan2018a} or  Leray  transform \cite{Lehoucq2015}. \modif{Fiber pattern anisotropy has been the subject of particular attention \cite{Boudaoud2014}.}
 Fourier transform (Fig. \ref{im:diffTF}) \modif{(FT)} has already been used to determine the anisotropy of \modif{fibrous-like} intra-cellular myosin distribution \cite{Bosveld2012}. 
 Fourier, Hilbert, and wavelet analysis are common in image analysis, with comparable performances  when tested on common benchmarks  \cite{Bruns2004,Bruns2005}. 
 One of the advantages of Fourier transform, beyond its simplicity, is that its amplitude (as opposed to its phase) is insensitive to small displacements of images; hence the Fourier amplitude measured on successive images, images from different regions, or images from different experiments can be averaged \cite{Bosveld2012}.

Here, we implement a Fourier \modif{transform-based} pipeline \modif{which, in addition to all above classical applications for pattern anisotropy quantifications, has specific advantages for the quantification of cellular patterns. It can determine the coarse grained} 
cell shape anisotropy in subregions of the whole image, resulting in cell shape anisotropy and orientation maps. 
Whenever \modif{the image quality is sufficient}, it can determine the cell size too. 
\modif{Note that it extracts the anisotropy and size of the averaged cell shape over a subregion ({\it not} the average of many individual cell anisotropies and sizes).} \modif{Whenever it is known, or it can be reasonably assumed, that in the rest state the cell shape is isotropic, the cell shape anisotropy in the current state measures the cell strain deviator (see Appendix \ref{sec:def}); similarly, if the cell size in the rest state is known, the current cell size measures the cell strain trace. These two measurements are fundamental for determining the mechanical state of the tissue.}
 We validate the pipeline with two already segmented images, in \modif{a} Drosophila pupa and in \modif{a} chicken embryo, and discuss its advantages.

\begin{figure}[h]
\includegraphics[scale = 0.75]{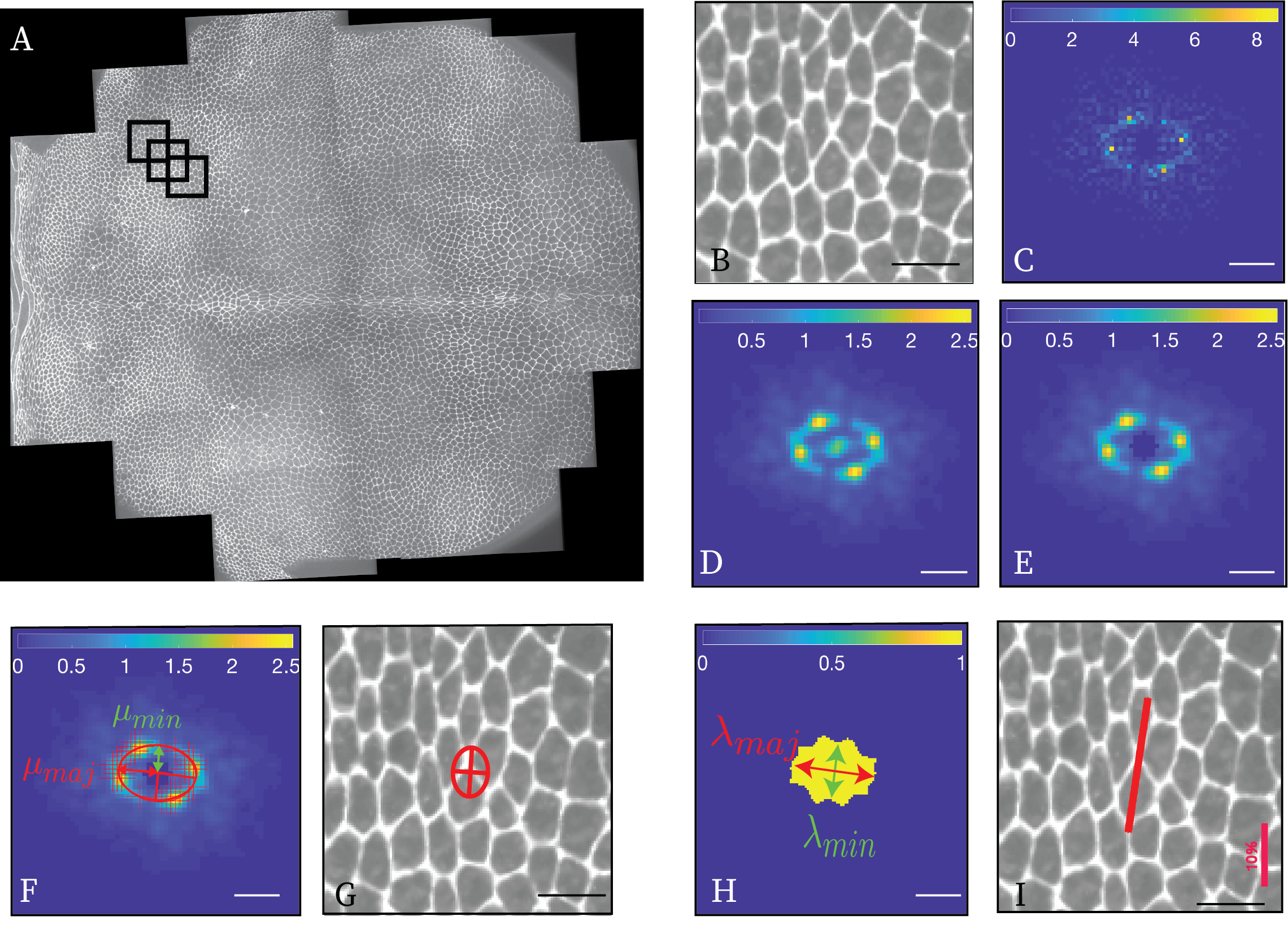}
\caption{Fourier transform\modif{-based} pipeline. $(A)$ Whole segmented image in original grey levels, same data as Fig.~\ref{fig:pup}A,B \cite{Guirao2015}. Boxes are examples of subwindows, showing their overlap.
$(B)$ Image in a subwindow, $(C)$ its power spectrum density, $(D)$ \modif{same} after a Gaussian blur of \modif{standard deviation 1.2},
$(E)$ \modif{same} after suppression of low frequencies (dark zone in the center). 
$(F)$ An ellipse is fitted \modif{to} the ring; its axis orientations and lengths $(\mu_{min},\mu_{maj})$  in the Fourier space define an ellipse with the same axis orientations and inverse axis lengths in the real space. $(G)$ \modif{In the real space} the ellipse size corresponds to the average cell size.
$(H)$ \modif{Thresholding the image, filling} the ellipse and measuring its FT inertia matrix \modif{determines the pattern anisotropy}, quantified by the ratio of ellipse axes $\lambda_{min}/\lambda_{maj}$ in the Fourier space, and the direction $\theta$ of the ellipse axes. \modif{Returning to} the real space $(I)$, the pattern anisotropy is determined: its magnitude is represented by the bar size, and its direction  is the direction of the bar. For $(C,D,E,F,H)$ \modif{white} scale bar is 0.1 $\mu$m$^{-1}$, for $(B,G,I$) \modif{black} scale bar is 10 $\mu$m, for $(I)$ \modif{red} scale bar is 10 $\%$ of elongation.}
\label{im:imMethod}
\end{figure}


\section{Pipeline}
\label{sec:pipeline}

The Fourier transform (FT) of an ordered periodic pattern has peaks (Fig. \ref{im:diffTF}A-C); the peak positions and intensities reveal the spatial periods and orientations present in the image. 
\modif{On the other hand,} the FT of a disordered anisotropic pattern (such as fibers) is a single broad \modif{peak} centered on zero frequency (Fig. \ref{im:diffTF}D-F); the distribution width reveals the range of length scales present in the image, and the distribution anisotropy reveals the fiber anisotropy.

An epithelial tissue pattern (Fig. \ref{im:imMethod}A) is aperiodic and usually lies in-between \modif{these} extremes.
The analysis is performed on overlapping subimages  (Fig.~\ref{im:imMethod}A). The window size is chosen as a trade-off between the signal to noise ratio and the spatial resolution required to answer the question under consideration.  
Fourier transform \modif{uses a periodic image, but in practice opposite borders of a window are different}. This causes artifacts - such as a cross on the FT spectra - that is avoided by a periodic plus smooth image decomposition \cite{Moisan2011}.
The FT phase provides information on the cell junction positions within each subimage, which is not of interest here. We keep only the FT amplitude.
In Fourier space, we represent the FT power spectrum density, with the  zero frequency at the center  (Fig.~\ref{im:imMethod}C). 
We perform a time average over successive images; their number is chosen as a trade-off between the signal to noise ratio and time resolution required to answer the question under consideration.
When the experiment is repeated, we average the FT spectrum of the different  available samples (``ensemble average"); as opposed to space and time averages, ensemble averaging has only advantages in terms of signal to noise ratio. 

The resulting power spectrum density is smoothed with a Gaussian blur (Fig.~\ref{im:imMethod}D). Low spatial frequencies, corresponding to lengthscales much larger than a cell size, are removed (Fig.~\ref{im:imMethod}E). 
The FT anisotropy reflects the pattern anisotropy; the FT itself is a  blurred ring (Fig.~\ref{im:imMethod}C-E), more or less \modif{resolved depending on} the initial image quality, and cell area variance. 
 This enables \modif{the two following} \modif{possibilities for the analysis of cell anisotropy}.

\modif{The first method, called the ``FT ellipse ring fit", also yields access to cell size. It applies to} a cellular pattern with \modif{disorderd} cell-cell junction orientation\modif{,} a small variance in area and a good image contrast, the FT is a well  \modif{resolved} elliptic ring  which can be fitted by an ellipse (Fig.~\ref{im:imMethod}F). 
Its axis sizes in Fourier space are $(\mu_{maj}$, $\mu_{min})$. They yield, back in real space, the ellipse axes sizes which describe the average cell properties within the subimage (Fig.~\ref{im:imMethod}G): $L_{maj} = \frac{2m}{\mu_{min}}$, $ L_{min} = \frac{2m}{\mu_{maj}}  $; here $m$ is the size of the FT image in pixels. To ensure the link with the real absolute size, $L_{maj}$ \modif{and} $L_{min}$ have to be multiplied by the pixel size. The angle between the $x$-axis and major axis is $\theta$ in real space and $\theta + \pi/2$ in Fourier space.

The second method\modif{, called the ``FT inertia matrix", is} more general because it applies even if the FT ellipse ring is \modif{ill-resolved}, as in Fig. \ref{im:diffTF}F. 
From Fig.~\ref{im:imMethod}E, \modif{we keep a percentile $p$ of the image pixels which are the brightest (hereafter called ``proportion" for short), to threshold the spectrum.}
A morphological closing is then performed to \modif{remove} the gaps between points (Fig.~\ref{im:imMethod}H). 
\modif{The resulting binarized pattern defines a} 
 \modif{filled} ellipse with a correct aspect ratio. Then, the inertia matrix (see Appendix \ref{sec:def}) of the \modif{filled} ellipse is computed and yields a major ($\lambda_{maj}$) and minor axis ($\lambda_{min}$). \modif{Returning to} the real space, the ellipse axes $L_{maj} = \frac{2m}{\lambda_{min}}$  and $L_{min} = \frac{2m}{\lambda_{maj}}$ define anisotropy.  Again, the angles of eigenvectors, $\theta$ and $\theta + \pi/2$, are the same in Fourier and real spaces. Note that here $L_{min}$ and $L_{maj}$ have no meaning in term\modif{s} of absolute cell size, as they are entirely dependent on the \modif{proportion} parameter. However, they reflect the pattern anisotropy, as we will now discuss.

There are several families of acceptable definitions of internal strain \cite{Bagi2006}. Among them, one contains an infinity of acceptable definitions that are functions of $L_{maj},L_{min}$ \cite{Farahani2000}. We choose the ``true" strain that was first introduced in the engineering field to describe large strains \cite{Hencky1931}. Using the true strain formalism the anisotropic part of the cell strain is defined as a matrix with the same eigenvectors as the FT and with eigenvalues $\pm \frac{1}{2}\log{\frac{L_{maj}}{L_{min}}}$  (see Appendix \ref{sec:def}). The absolute value of this amplitude (or its linearized approximations, if the strain is small, see Appendix \ref{sec:def}) is used as a measure of anisotropy, \modif{which} we represent as a bar in the direction $\theta$ (Fig.~\ref{im:imMethod}I).
The results are sensitive to the \modif{proportion $p$} of pixels kept for thresholding. However, a reasonable range of values \modif{of $p$ allows for a robust determination of}  anisotropies (see Appendix \ref{sec:paramrange}).

Altogether, the parameters \modif{which must be adjusted} for both methods are: window size and overlap, time average, Gaussian blur size, low cut-off for spatial frequencies; in addition, for FT ellipse ring fit: number of fit points; and for FT inertia matrix:  \modif{proportion} for thresholding, dilation-erosion size.
\modif{The code is available on \href{https://github.com/mdurande/coarse-grained-anisotropy-and-size-using-FFT}{Github} \cite{Durande:2019}}. It is user friendly and optimised to reduce the time it takes to manually adjust the parameters, typically 5 minutes at the beginning and 1 minute when the user is trained. Once these parameters are adjusted for a first image, they can be re-used for all similar images of the same series.
  
\section{Results}
\label{sec:results}
  
\subsection{Precision on cell size determination}
\label{sec:resolution}
  
\begin{figure}[h]
\includegraphics[scale = 0.9]{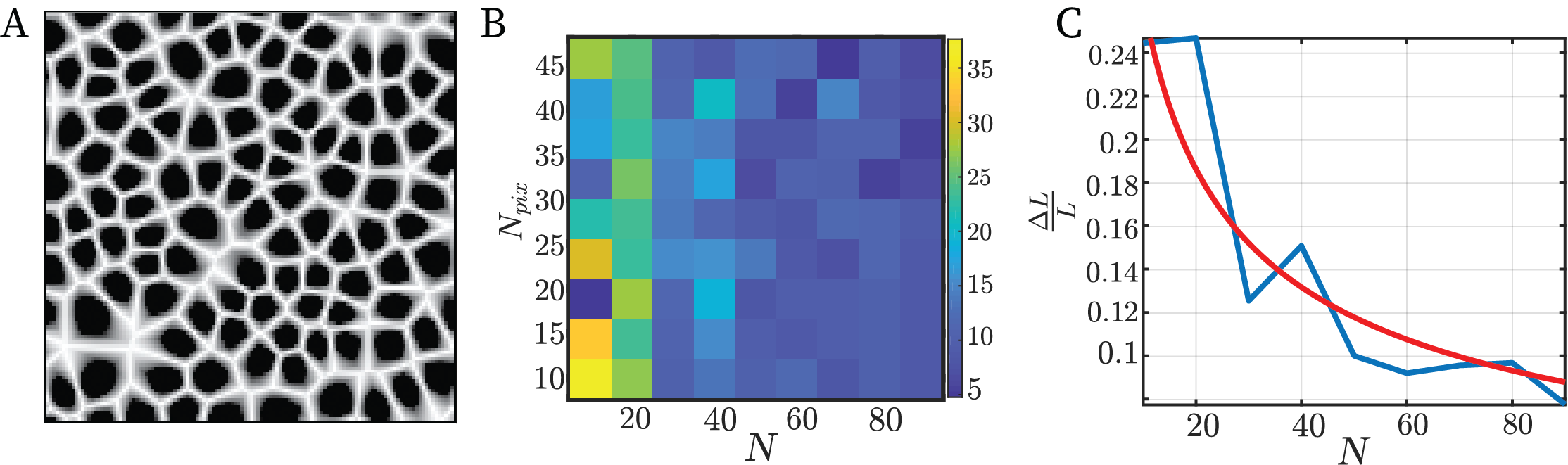}
\caption{Precision on cell size determination. \textit{(A)} Example of an artificial cellular pattern.  Window size 128 pixels, at least 8 pixels between seeds, packing fraction 0.5,  $N=80$ cells,  $N_{pix}=158$ pixels per cell in average. \textit{(B)} Relative error on the cell size: $\frac{\Delta L}{L}$ \modif{in percentage}, where $L$ is the average radius of the cells in the image vs number $N$ of cells in the image and average number $N_{pix}$  of pixels per cells. Data from 1260 images (126 series of 10 repeats). 
Each square is an average with a minimum of 10 images. \textit{(C)} Blue: relative error vs $N$ for all images, ie: averaged on $N_{pix}$. Red: best fit  by a $N^{-1/2}$ law, prefactor 0.83.}
\label{im:fakecells}
\end{figure}

 To test the precision on cell size determination, we first \modif{run the pipeline} on a set of artificial cellular patterns. 
Each image is created on a square of side $L_{pix}$ pixels by  sequentially placing $N$ seeds  at random points, with a minimum distance between them. Their Voronoi diagram is created, and the cell-cell junctions are thickened to reach a prescribed packing fraction (Fig.~\ref{im:fakecells}A). We measure the number of pixels per cell and the average cell size on the pattern. We then apply our pipeline and compute its error in cell size determination. This test is repeated on a series of 10 images with the same parameters (minimum distance between seeds, and packing fraction). Then the parameters are varied to generate a set of \modif{126} different series.


 Since we measure the cell size $L$ from a peak in the FT, we expect the peak position in Fourier space to be around $1/L$. The precision in peak position determination is of \modif{the} order of one pixel in Fourier space, ie: $\frac{1}{L\sqrt{N}}$ \modif{back in real space}. The relative error on $L$ is thus of order $ N^{-1/2}$, independently of the number of pixels per cells $N_{pix}$. 
 This is consistent with the results of our tests, where the value of $N_{pix}$ has no effect  as soon as it is larger than 20 (Fig.~\ref{im:fakecells}B) and the value of \modif{$\Delta L/L$} is of \modif{the} order of 0.83~$N^{-1/2}$ (Fig.~\ref{im:fakecells}B,C). \modif{Note that it would be possible to increase the resolution by padding the image - adding zeros around the picture  \cite{Hilbert:2013}. This simple process allows the pixel size in Fourier \modif{space} to be changed, and thus gives access to different ranges of frequency: it can \modif{improve the  Fourier transform resolution and allow a sub-pixel accuracy to be reached back in real space.} It is not used in the present article nor in the online code.}

\subsection{\modif{Validation of} cell size and anisotropy determination}

To \modif{validate} the cell size and anisotropy determination methods, we \modif{run the pipeline on} an image  (Fig.~\ref{im:imMethod}A) \modif{whose} segmentation (Fig.~\ref{fig:pup}A) \modif{quality makes it a gold standard} \cite{Guirao2015}. 
  The FT calculation has been performed \modif{in Matlab} on a OSX with an Intel Core i7 processor at 2.2 GHz clock frequency. It takes about 60 minutes for the computation of the anisotropy part alone with the inertia matrix method, about 40 seconds for the computation of the size alone \modif{with the ellipse ring fit method}. 


\begin{figure}[h!]
\includegraphics[scale = 0.75]{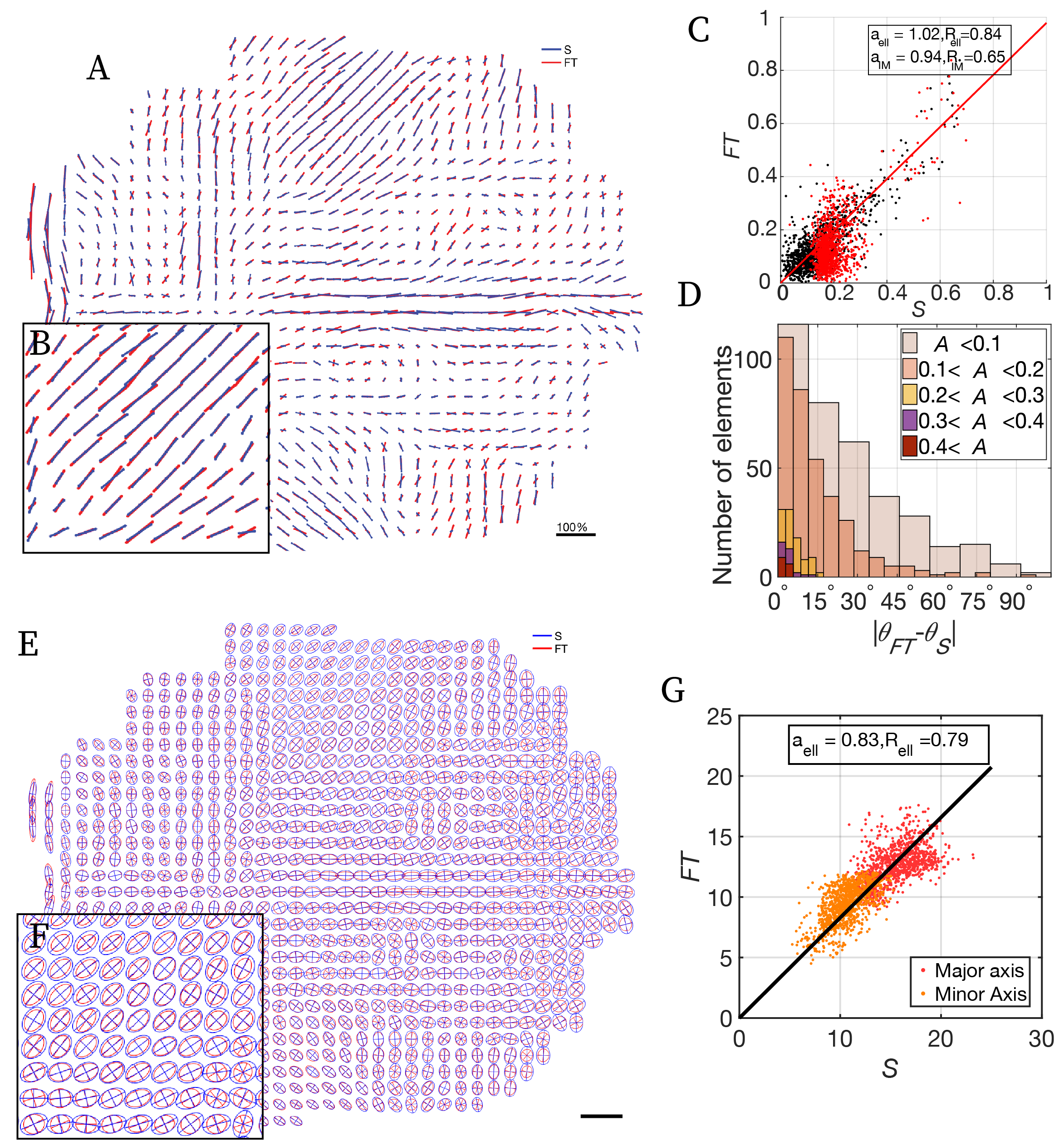}
\caption{Test of cell anisotropy and size measurements.
$(A)$ Map of cell anisotropies on the image in Fig.~\ref{fig:pup}A. There are 1221 boxes of 128 $\times$ 128 pixels with a 50$\%$ overlap. 
 Blue bars: from segmentation, data of Fig.~\ref{fig:pup}A. Red bars:  from FT analysis computed with the inertia matrix.
 $(B)$ \modif{Close-up} to a region of $(A)$.
  $(C)$ \modif{Anisotropy} measurements using FT ellipse fit ring (black) and FT inertia matrix (red) vs measurements using segmentation. Each point corresponds to a box. 
  The slopes of the fit \modif{are} 1.02 ($R=0.84$)  for FT ellipse fit ring and 0.94 ($R=0.65$) for FT inertia matrix.
$(D)$ Histogram of the difference between angles from FT with the inertia matrix method, and from segmentation. The color codes for the anisotropy amplitude $A$. 
$(E)$ Map of cell ellipses, representing cell sizes and anisotropies.
 Blue: \modif{results} from segmentation, data of Fig.~\ref{fig:pup}A. Red: \modif{results from FT ellipse ring fit method, plotted as ellipses in real space}. Scale bar is 50 $\mu$m.
 $(F)$ Close up to a region of $(E)$.
  $(G)$ \modif{Major axis (red) and minor axis (orange)} measurements using FT ellipse fit vs measurements using segmentation. Each point corresponds to a box;
 the slope of the fit is 0.83 ($R=0.79$).
}
\label{im:validampi}
\end{figure}

The cell anisotropy measurements using FT inertia matrix methods correlate with the segmentation measurements, qualitatively (Fig.~\ref{im:validampi}A,B) and quantitatively  (Fig.~\ref{im:validampi}C,D) in amplitude and orientation. The cell size and anisotropy \modif{(amplitude and orientation)} measurements using the FT ellipse ring fit correlate well with the segmentation measurements, qualitatively  (Fig.~\ref{im:validampi}E,F) and quantitatively  (Fig.~\ref{im:validampi}C,G). As expected, the anisotropy orientation measurement is better when the anisotropy amplitude is larger; at small anisotropies the  FT ellipse ring fit performs better than the FT inertia matrix (Fig.~\ref{im:validampi}C).

\subsection{Measurements on \modif{a} large dataset}
  
\begin{figure}[h]
\includegraphics[scale=1]{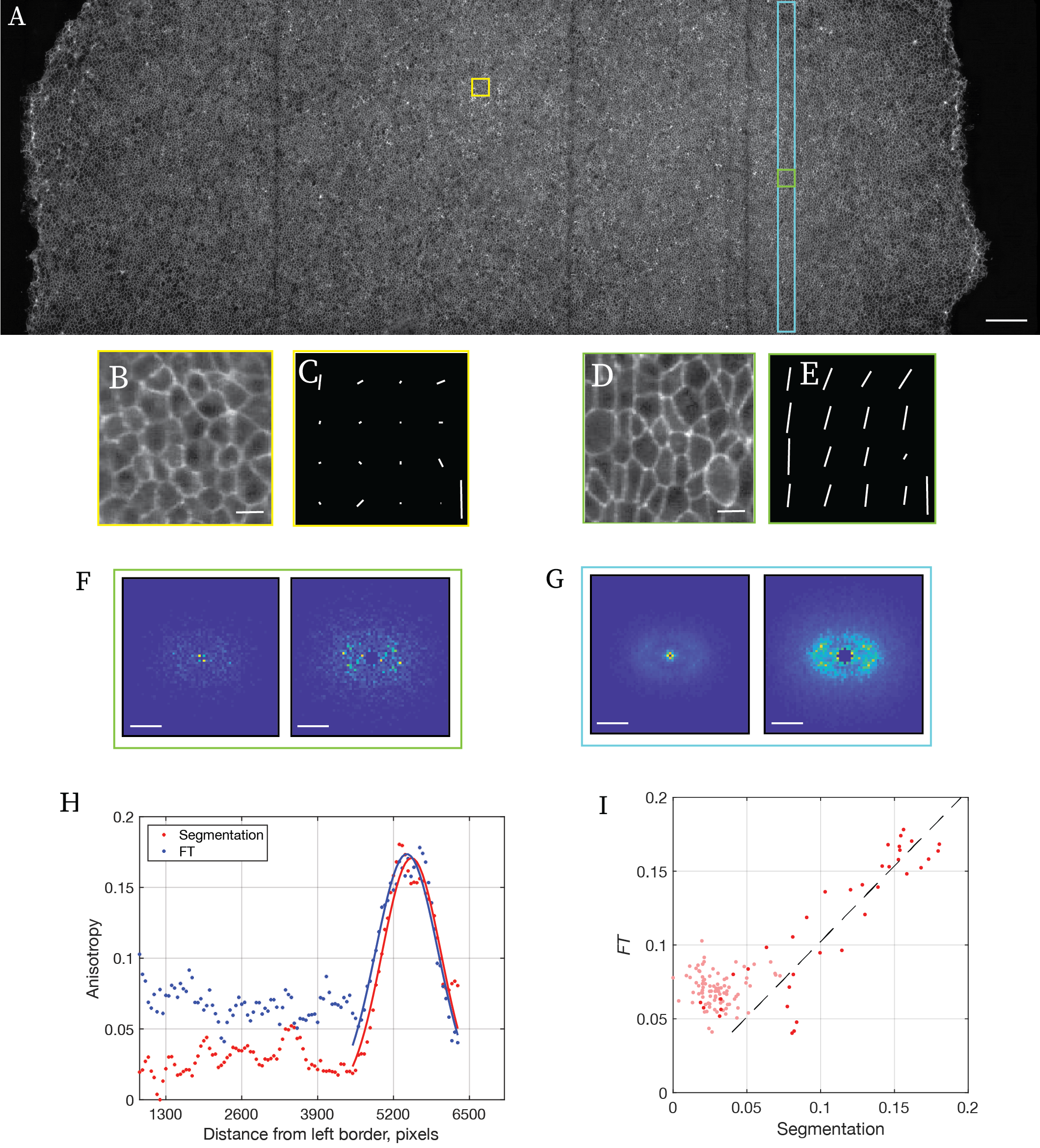}
\caption{\modif{Large cell number} anisotropy measurements.
$(A)$
Light sheet microscopy image of \modif{a} chicken embryo at cellular resolution  \cite{Rozbicki2015}. Scale bar is 200 $\mu$m. The field of view contains \modif{of order of $10^5$} cells. Anterioposterior (AP) axis is horizontal, with anterior on the left, posterior on the right. The two \modif{framed squares} are  200 $\times$ 200 pixels ($\sim$100 cells).
$(B,C)$
Close up of yellow \modif{framed square} in $A$, and corresponding anisotropy measured by FT inertia matrix \modif{at several positions}.
$(D,E)$ Same as $B$, $C$ for green \modif{framed square} in  $A$, in the sickle region. For $(B,D)$ scale bar is 20 $\mu$m; for $(C,E)$ scale bar is 10 $\%$ elongation.
$(F)$
 Left: raw Fourier spectrum of the green \modif{framed square} in $A$. Right: Same after removing small spatial frequencies and adapting the color scale. White scale bar is $10^{-1} \mu$m$^{-1}$. 
$(G)$ Same as $F$, averaged over the whole blue \modif{framed rectangle} in $A$, perpendicular to the AP axis.
$(H)$ Anisotropy vs position along the AP axis. Averages are performed perpendicularly to the AP axis, as in $G$, and the anisotropy computed after averaging. 
Red: measurements using segmentation; blue: measurements using  FT inertia matrix; \modif{lines: Gaussian fits to the region between 4500 and 6300 pixels from left border.}
$(I)$
Measurements using  FT inertia matrix vs measurements using segmentation. Dashed line: linear fit  \modif{to the same data as in $H$}  
\modif{(in red)}, slope  1.0\modif{2} ($R=0.\modif{89}$).  
}
\label{Fig:chicken}
\end{figure}

We now test the FT analysis on a case where the cell number is particularly large. 
Data come from chicken morphogenesis, more precisely from a study of cell flows during primitive streak formation, estimating the relative contributions of cell shape changes and cell neighbour rearrangements  \cite{Rozbicki2015}.
Each image contains hundreds of thousands of cells (Fig.~\ref{Fig:chicken}A). 
Altogether, taking into account wild-type and mutant conditions, hundreds of movies have been acquired, each with hundreds of images, resulting in several billion cells. 
The image quality and contrast  are good enough for segmentation, but the cell number is too large and segmentation has been performed only on a subset of images.


Most boxes display clearly isotropic cell shapes (Fig.~\ref{Fig:chicken}B), as quantitatively confirmed by their anisotropy (Fig.~\ref{Fig:chicken}C). However, \modif{several} boxes contain cells clearly displaying \modif{a} shape anisotropy (Fig.~\ref{Fig:chicken}D), as again quantitatively confirmed  (Fig.~\ref{Fig:chicken}E). These anisotropic cells are all located in a band, perpendicular to the anterioposterior (AP) axis, the so-called sickle region undergoing an extension.
The \modif{measurement which is sought here} is the  \modif{position and width of this extension} region.  This does not require segmentation, and FT analysis seems appropriate. 

Taking advantage of the expected band structure, we average the FT spectrum (Fig.~\ref{Fig:chicken}F) over boxes in a rectangle perpendicular to the AP axis, strongly improving the signal to noise ratio (Fig.~\ref{Fig:chicken}G):
the ellipse ring becomes visible. We have the choice between both methods and choose here the FT inertia matrix, more robust than the FT ellipse ring fit to variations in image quality and contrast \modif{and sufficient to answer the questions under consideration}.
By thresholding, calculating the inertia matrix and diagonalizing it, we measure the anisotropy of the average FT (not the average of FT anisotropy).
The calculation has been performed on a OSX with an Intel Core i7 processor at 2.2 GHz clock frequency. It takes 3 minutes in \modif{Matlab} to compute the anisotropy part alone with the inertia matrix method.

FT provides the whole profile of anisotropy along the $x$ axis, in  good qualitative agreement with the segmentation.
Quantitatively, for boxes with anisotropy at least equal to 0.08, the agreement between FT and segmentation results is very good (Fig.~\ref{Fig:chicken}H,I)\modif{: for data obtained with segmentation, the Gaussian fit peak position is at 5500 $\pm$ 11 $\mu$m and its \modif{standard deviation} is $\sigma$ = 358 $\pm$ 23 $\mu$m, while for the FT inertia matrix data, the peak position is at 5450 $\pm$ 10 $\mu$m and  $\sigma$ = 377 $\pm$ 20 $\mu$m.}

The FT clearly \modif{reveals} the existence \modif{and estimates the} position and width of the band, in good qualitative agreement with the segmentation (Fig.~\ref{Fig:chicken}H).

\section{Conclusion}

In summary, we \modif{present} a Fourier \modif{transform-based} analysis pipeline to measure the  coarse grained field of pattern anisotropies.  It applies \modif{in particular} to fixed or live,  fluorescent and phase contrast images of epithelial tissues\modif{, in which it characterizes the coarse grained cell anisotropy. One variant, the ``inertia matrix" method, applies even when the image quality is low. The other variant, the ``ellipse ring" method, yields also the coarse grained  cell size}\modif{, and performs better at small anisotropies.}
 
 After a user-friendly manual tuning of a few parameters, it can automatically handle tens of thousands of cells per second. We  successfully validated it against segmentation based measurements. It is robust against \modif{defects} in image contrast,  heterogeneities in cell sizes and orientations\modif{, choice of parameters}. It is adequate  to improve the signal to noise ratio using space, time and/or ensemble averages; the latter are averages over samples and do not deteriorate the time or space resolution. 

Extensions of applications could include ordered tissues, non-living cellular materials such as liquid foams, real-time analysis of live movies \modif{and/or} tri-dimensional tissues.
 \modif{Moreover, \modif{assuming that in the rest state the coarse grained cell shape is isotropic,}  the pipeline can also yield access to the cell strain deviator, a mechanical measure important to characterize a tissue}\modif{, as shown in a companion paper \cite{Tlili2018b}.}
\begin{acknowledgments}

We warmly thank C.J. Weijer for providing chicken embryo images and for critical reading of the manuscript. We also thank F. Bosveld, S. Rigaud and Y. Bella\"{\i}che for their role in acquiring Drosophila pupa data.

\end{acknowledgments}
\bibliography{bilbioTF}

\begin{thebibliography}{34}%
\makeatletter
\providecommand \@ifxundefined [1]{%
 \@ifx{#1\undefined}
}%
\providecommand \@ifnum [1]{%
 \ifnum #1\expandafter \@firstoftwo
 \else \expandafter \@secondoftwo
 \fi
}%
\providecommand \@ifx [1]{%
 \ifx #1\expandafter \@firstoftwo
 \else \expandafter \@secondoftwo
 \fi
}%
\providecommand \natexlab [1]{#1}%
\providecommand \enquote  [1]{``#1''}%
\providecommand \bibnamefont  [1]{#1}%
\providecommand \bibfnamefont [1]{#1}%
\providecommand \citenamefont [1]{#1}%
\providecommand \href@noop [0]{\@secondoftwo}%
\providecommand \href [0]{\begingroup \@sanitize@url \@href}%
\providecommand \@href[1]{\@@startlink{#1}\@@href}%
\providecommand \@@href[1]{\endgroup#1\@@endlink}%
\providecommand \@sanitize@url [0]{\catcode `\\12\catcode `\$12\catcode
  `\&12\catcode `\#12\catcode `\^12\catcode `\_12\catcode `\%12\relax}%
\providecommand \@@startlink[1]{}%
\providecommand \@@endlink[0]{}%
\providecommand \url  [0]{\begingroup\@sanitize@url \@url }%
\providecommand \@url [1]{\endgroup\@href {#1}{\urlprefix }}%
\providecommand \urlprefix  [0]{URL }%
\providecommand \Eprint [0]{\href }%
\providecommand \doibase [0]{http://dx.doi.org/}%
\providecommand \selectlanguage [0]{\@gobble}%
\providecommand \bibinfo  [0]{\@secondoftwo}%
\providecommand \bibfield  [0]{\@secondoftwo}%
\providecommand \translation [1]{[#1]}%
\providecommand \BibitemOpen [0]{}%
\providecommand \bibitemStop [0]{}%
\providecommand \bibitemNoStop [0]{.\EOS\space}%
\providecommand \EOS [0]{\spacefactor3000\relax}%
\providecommand \BibitemShut  [1]{\csname bibitem#1\endcsname}%
\let\auto@bib@innerbib\@empty
\bibitem [{\citenamefont {Heisenberg}\ and\ \citenamefont
  {Bella{\"{i}}che}(2013)}]{Heisenberg2013}%
  \BibitemOpen
  \bibfield  {author} {\bibinfo {author} {\bibfnamefont {C.-P.}\ \bibnamefont
  {Heisenberg}}\ and\ \bibinfo {author} {\bibfnamefont {Y.}~\bibnamefont
  {Bella{\"{i}}che}},\ }\href {\doibase 10.1016/j.cell.2013.05.008} {\bibfield
  {journal} {\bibinfo  {journal} {Cell}\ }\textbf {\bibinfo {volume} {153}},\
  \bibinfo {pages} {948} (\bibinfo {year} {2013})}\BibitemShut {NoStop}%
\bibitem [{\citenamefont {Zallen}\ and\ \citenamefont
  {Zallen}(2004)}]{Zallen2004}%
  \BibitemOpen
  \bibfield  {author} {\bibinfo {author} {\bibfnamefont {J.~A.}\ \bibnamefont
  {Zallen}}\ and\ \bibinfo {author} {\bibfnamefont {R.}~\bibnamefont
  {Zallen}},\ }\href {\doibase 10.1088/0953-8984/16/44/005} {\bibfield
  {journal} {\bibinfo  {journal} {J. Phys. Condens. Matter}\ }\textbf {\bibinfo
  {volume} {16}},\ \bibinfo {pages} {5073} (\bibinfo {year}
  {2004})}\BibitemShut {NoStop}%
\bibitem [{\citenamefont {Classen}\ \emph {et~al.}(2005)\citenamefont
  {Classen}, \citenamefont {Anderson}, \citenamefont {Marois},\ and\
  \citenamefont {Eaton}}]{Classen2005}%
  \BibitemOpen
  \bibfield  {author} {\bibinfo {author} {\bibfnamefont {A.-K.}\ \bibnamefont
  {Classen}}, \bibinfo {author} {\bibfnamefont {K.~I.}\ \bibnamefont
  {Anderson}}, \bibinfo {author} {\bibfnamefont {E.}~\bibnamefont {Marois}}, \
  and\ \bibinfo {author} {\bibfnamefont {S.}~\bibnamefont {Eaton}},\ }\href
  {\doibase 10.1016/J.DEVCEL.2005.10.016} {\bibfield  {journal} {\bibinfo
  {journal} {Dev. Cell}\ }\textbf {\bibinfo {volume} {9}},\ \bibinfo {pages}
  {805} (\bibinfo {year} {2005})}\BibitemShut {NoStop}%
\bibitem [{\citenamefont {Hayashi}\ and\ \citenamefont
  {Carthew}(2004)}]{Hayashi2004}%
  \BibitemOpen
  \bibfield  {author} {\bibinfo {author} {\bibfnamefont {T.}~\bibnamefont
  {Hayashi}}\ and\ \bibinfo {author} {\bibfnamefont {R.~W.}\ \bibnamefont
  {Carthew}},\ }\href {\doibase 10.1038/nature02952} {\bibfield  {journal}
  {\bibinfo  {journal} {Nature}\ }\textbf {\bibinfo {volume} {431}},\ \bibinfo
  {pages} {647} (\bibinfo {year} {2004})}\BibitemShut {NoStop}%
\bibitem [{\citenamefont {K{\"{a}}fer}\ \emph {et~al.}(2007)\citenamefont
  {K{\"{a}}fer}, \citenamefont {Hayashi}, \citenamefont {Mar{\'{e}}e},
  \citenamefont {Carthew},\ and\ \citenamefont {Graner}}]{Kafer2007}%
  \BibitemOpen
  \bibfield  {author} {\bibinfo {author} {\bibfnamefont {J.}~\bibnamefont
  {K{\"{a}}fer}}, \bibinfo {author} {\bibfnamefont {T.}~\bibnamefont
  {Hayashi}}, \bibinfo {author} {\bibfnamefont {A.~F.~M.}\ \bibnamefont
  {Mar{\'{e}}e}}, \bibinfo {author} {\bibfnamefont {R.~W.}\ \bibnamefont
  {Carthew}}, \ and\ \bibinfo {author} {\bibfnamefont {F.}~\bibnamefont
  {Graner}},\ }\href {\doibase 10.1073/pnas.0704235104} {\bibfield  {journal}
  {\bibinfo  {journal} {Proc. Natl. Acad. Sci. U. S. A.}\ }\textbf {\bibinfo
  {volume} {104}},\ \bibinfo {pages} {18549} (\bibinfo {year}
  {2007})}\BibitemShut {NoStop}%
\bibitem [{\citenamefont {Hilgenfeldt}\ \emph {et~al.}(2008)\citenamefont
  {Hilgenfeldt}, \citenamefont {Erisken},\ and\ \citenamefont
  {Carthew}}]{Hilgenfeldt2008}%
  \BibitemOpen
  \bibfield  {author} {\bibinfo {author} {\bibfnamefont {S.}~\bibnamefont
  {Hilgenfeldt}}, \bibinfo {author} {\bibfnamefont {S.}~\bibnamefont
  {Erisken}}, \ and\ \bibinfo {author} {\bibfnamefont {R.~W.}\ \bibnamefont
  {Carthew}},\ }\href {\doibase 10.1073/pnas.0711077105} {\bibfield  {journal}
  {\bibinfo  {journal} {Proc. Natl. Acad. Sci. U. S. A.}\ }\textbf {\bibinfo
  {volume} {105}},\ \bibinfo {pages} {907} (\bibinfo {year}
  {2008})}\BibitemShut {NoStop}%
\bibitem [{\citenamefont {Farhadifar}\ \emph {et~al.}(2007)\citenamefont
  {Farhadifar}, \citenamefont {R{\"{o}}per}, \citenamefont {Aigouy},
  \citenamefont {Eaton},\ and\ \citenamefont
  {J{\"{u}}licher}}]{Farhadifar2007}%
  \BibitemOpen
  \bibfield  {author} {\bibinfo {author} {\bibfnamefont {R.}~\bibnamefont
  {Farhadifar}}, \bibinfo {author} {\bibfnamefont {J.-C.}\ \bibnamefont
  {R{\"{o}}per}}, \bibinfo {author} {\bibfnamefont {B.}~\bibnamefont {Aigouy}},
  \bibinfo {author} {\bibfnamefont {S.}~\bibnamefont {Eaton}}, \ and\ \bibinfo
  {author} {\bibfnamefont {F.}~\bibnamefont {J{\"{u}}licher}},\ }\href
  {\doibase 10.1016/J.CUB.2007.11.049} {\bibfield  {journal} {\bibinfo
  {journal} {Curr. Biol.}\ }\textbf {\bibinfo {volume} {17}},\ \bibinfo {pages}
  {2095} (\bibinfo {year} {2007})}\BibitemShut {NoStop}%
\bibitem [{\citenamefont {Rauzi}\ \emph {et~al.}(2008)\citenamefont {Rauzi},
  \citenamefont {Verant}, \citenamefont {Lecuit},\ and\ \citenamefont
  {Lenne}}]{Rauzi2008}%
  \BibitemOpen
  \bibfield  {author} {\bibinfo {author} {\bibfnamefont {M.}~\bibnamefont
  {Rauzi}}, \bibinfo {author} {\bibfnamefont {P.}~\bibnamefont {Verant}},
  \bibinfo {author} {\bibfnamefont {T.}~\bibnamefont {Lecuit}}, \ and\ \bibinfo
  {author} {\bibfnamefont {P.-F.}\ \bibnamefont {Lenne}},\ }\href {\doibase
  10.1038/ncb1798} {\bibfield  {journal} {\bibinfo  {journal} {Nat. Cell
  Biol.}\ }\textbf {\bibinfo {volume} {10}},\ \bibinfo {pages} {1401} (\bibinfo
  {year} {2008})}\BibitemShut {NoStop}%
\bibitem [{\citenamefont {Blanchard}\ \emph {et~al.}(2009)\citenamefont
  {Blanchard}, \citenamefont {Kabla}, \citenamefont {Schultz}, \citenamefont
  {Butler}, \citenamefont {Sanson}, \citenamefont {Gorfinkiel}, \citenamefont
  {Mahadevan},\ and\ \citenamefont {Adams}}]{Blanchard2009}%
  \BibitemOpen
  \bibfield  {author} {\bibinfo {author} {\bibfnamefont {G.~B.}\ \bibnamefont
  {Blanchard}}, \bibinfo {author} {\bibfnamefont {A.~J.}\ \bibnamefont
  {Kabla}}, \bibinfo {author} {\bibfnamefont {N.~L.}\ \bibnamefont {Schultz}},
  \bibinfo {author} {\bibfnamefont {L.~C.}\ \bibnamefont {Butler}}, \bibinfo
  {author} {\bibfnamefont {B.}~\bibnamefont {Sanson}}, \bibinfo {author}
  {\bibfnamefont {N.}~\bibnamefont {Gorfinkiel}}, \bibinfo {author}
  {\bibfnamefont {L.}~\bibnamefont {Mahadevan}}, \ and\ \bibinfo {author}
  {\bibfnamefont {R.~J.}\ \bibnamefont {Adams}},\ }\href {\doibase
  10.1038/nmeth.1327} {\bibfield  {journal} {\bibinfo  {journal} {Nat.
  Methods}\ }\textbf {\bibinfo {volume} {6}},\ \bibinfo {pages} {458} (\bibinfo
  {year} {2009})}\BibitemShut {NoStop}%
\bibitem [{\citenamefont {Guirao}\ \emph {et~al.}(2015)\citenamefont {Guirao},
  \citenamefont {Rigaud}, \citenamefont {Bosveld}, \citenamefont {Bailles},
  \citenamefont {L{\'{o}}pez-Gay}, \citenamefont {Ishihara}, \citenamefont
  {Sugimura}, \citenamefont {Graner},\ and\ \citenamefont
  {Bella{\"{i}}che}}]{Guirao2015}%
  \BibitemOpen
  \bibfield  {author} {\bibinfo {author} {\bibfnamefont {B.}~\bibnamefont
  {Guirao}}, \bibinfo {author} {\bibfnamefont {S.~U.}\ \bibnamefont {Rigaud}},
  \bibinfo {author} {\bibfnamefont {F.}~\bibnamefont {Bosveld}}, \bibinfo
  {author} {\bibfnamefont {A.}~\bibnamefont {Bailles}}, \bibinfo {author}
  {\bibfnamefont {J.}~\bibnamefont {L{\'{o}}pez-Gay}}, \bibinfo {author}
  {\bibfnamefont {S.}~\bibnamefont {Ishihara}}, \bibinfo {author}
  {\bibfnamefont {K.}~\bibnamefont {Sugimura}}, \bibinfo {author}
  {\bibfnamefont {F.}~\bibnamefont {Graner}}, \ and\ \bibinfo {author}
  {\bibfnamefont {Y.}~\bibnamefont {Bella{\"{i}}che}},\ }\href {\doibase
  10.7554/eLife.08519} {\bibfield  {journal} {\bibinfo  {journal} {Elife}\
  }\textbf {\bibinfo {volume} {4}},\ \bibinfo {pages} {e08519} (\bibinfo {year}
  {2015})}\BibitemShut {NoStop}%
\bibitem [{\citenamefont {Etournay}\ \emph {et~al.}(2016)\citenamefont
  {Etournay}, \citenamefont {Merkel}, \citenamefont {Popovi{\'{c}}},
  \citenamefont {Brandl}, \citenamefont {Dye}, \citenamefont {Aigouy},
  \citenamefont {Salbreux}, \citenamefont {Eaton},\ and\ \citenamefont
  {J{\"{u}}licher}}]{Etournay2016}%
  \BibitemOpen
  \bibfield  {author} {\bibinfo {author} {\bibfnamefont {R.}~\bibnamefont
  {Etournay}}, \bibinfo {author} {\bibfnamefont {M.}~\bibnamefont {Merkel}},
  \bibinfo {author} {\bibfnamefont {M.}~\bibnamefont {Popovi{\'{c}}}}, \bibinfo
  {author} {\bibfnamefont {H.}~\bibnamefont {Brandl}}, \bibinfo {author}
  {\bibfnamefont {N.~A.}\ \bibnamefont {Dye}}, \bibinfo {author} {\bibfnamefont
  {B.}~\bibnamefont {Aigouy}}, \bibinfo {author} {\bibfnamefont
  {G.}~\bibnamefont {Salbreux}}, \bibinfo {author} {\bibfnamefont
  {S.}~\bibnamefont {Eaton}}, \ and\ \bibinfo {author} {\bibfnamefont
  {F.}~\bibnamefont {J{\"{u}}licher}},\ }\href {\doibase 10.7554/eLife.14334}
  {\bibfield  {journal} {\bibinfo  {journal} {Elife}\ }\textbf {\bibinfo
  {volume} {5}} (\bibinfo {year} {2016}),\ 10.7554/eLife.14334}\BibitemShut
  {NoStop}%
\bibitem [{\citenamefont {Faure}\ \emph {et~al.}(2016)\citenamefont {Faure},
  \citenamefont {Savy}, \citenamefont {Rizzi}, \citenamefont {Melani},
  \citenamefont {Sta{\v{s}}ov{\'{a}}}, \citenamefont {Fabr{\`{e}}ges},
  \citenamefont {{\v{S}}pir}, \citenamefont {Hammons}, \citenamefont
  {{\v{C}}{\'{u}}nderl{\'{i}}k}, \citenamefont {Recher}, \citenamefont
  {Lombardot}, \citenamefont {Duloquin}, \citenamefont {Colin}, \citenamefont
  {Koll{\'{a}}r}, \citenamefont {Desnoulez}, \citenamefont {Affaticati},
  \citenamefont {Maury}, \citenamefont {Boyreau}, \citenamefont {Nief},
  \citenamefont {Calvat}, \citenamefont {Vernier}, \citenamefont {Frain},
  \citenamefont {Lutfalla}, \citenamefont {Kergosien}, \citenamefont {Suret},
  \citenamefont {Reme{\v{s}}{\'{i}}kov{\'{a}}}, \citenamefont {Doursat},
  \citenamefont {Sarti}, \citenamefont {Mikula}, \citenamefont
  {Peyri{\'{e}}ras},\ and\ \citenamefont {Bourgine}}]{Faure2016}%
  \BibitemOpen
  \bibfield  {author} {\bibinfo {author} {\bibfnamefont {E.}~\bibnamefont
  {Faure}}, \bibinfo {author} {\bibfnamefont {T.}~\bibnamefont {Savy}},
  \bibinfo {author} {\bibfnamefont {B.}~\bibnamefont {Rizzi}}, \bibinfo
  {author} {\bibfnamefont {C.}~\bibnamefont {Melani}}, \bibinfo {author}
  {\bibfnamefont {O.}~\bibnamefont {Sta{\v{s}}ov{\'{a}}}}, \bibinfo {author}
  {\bibfnamefont {D.}~\bibnamefont {Fabr{\`{e}}ges}}, \bibinfo {author}
  {\bibfnamefont {R.}~\bibnamefont {{\v{S}}pir}}, \bibinfo {author}
  {\bibfnamefont {M.}~\bibnamefont {Hammons}}, \bibinfo {author} {\bibfnamefont
  {R.}~\bibnamefont {{\v{C}}{\'{u}}nderl{\'{i}}k}}, \bibinfo {author}
  {\bibfnamefont {G.}~\bibnamefont {Recher}}, \bibinfo {author} {\bibfnamefont
  {B.}~\bibnamefont {Lombardot}}, \bibinfo {author} {\bibfnamefont
  {L.}~\bibnamefont {Duloquin}}, \bibinfo {author} {\bibfnamefont
  {I.}~\bibnamefont {Colin}}, \bibinfo {author} {\bibfnamefont
  {J.}~\bibnamefont {Koll{\'{a}}r}}, \bibinfo {author} {\bibfnamefont
  {S.}~\bibnamefont {Desnoulez}}, \bibinfo {author} {\bibfnamefont
  {P.}~\bibnamefont {Affaticati}}, \bibinfo {author} {\bibfnamefont
  {B.}~\bibnamefont {Maury}}, \bibinfo {author} {\bibfnamefont
  {A.}~\bibnamefont {Boyreau}}, \bibinfo {author} {\bibfnamefont {J.-Y.}\
  \bibnamefont {Nief}}, \bibinfo {author} {\bibfnamefont {P.}~\bibnamefont
  {Calvat}}, \bibinfo {author} {\bibfnamefont {P.}~\bibnamefont {Vernier}},
  \bibinfo {author} {\bibfnamefont {M.}~\bibnamefont {Frain}}, \bibinfo
  {author} {\bibfnamefont {G.}~\bibnamefont {Lutfalla}}, \bibinfo {author}
  {\bibfnamefont {Y.}~\bibnamefont {Kergosien}}, \bibinfo {author}
  {\bibfnamefont {P.}~\bibnamefont {Suret}}, \bibinfo {author} {\bibfnamefont
  {M.}~\bibnamefont {Reme{\v{s}}{\'{i}}kov{\'{a}}}}, \bibinfo {author}
  {\bibfnamefont {R.}~\bibnamefont {Doursat}}, \bibinfo {author} {\bibfnamefont
  {A.}~\bibnamefont {Sarti}}, \bibinfo {author} {\bibfnamefont
  {K.}~\bibnamefont {Mikula}}, \bibinfo {author} {\bibfnamefont
  {N.}~\bibnamefont {Peyri{\'{e}}ras}}, \ and\ \bibinfo {author} {\bibfnamefont
  {P.}~\bibnamefont {Bourgine}},\ }\href {\doibase 10.1038/ncomms9674}
  {\bibfield  {journal} {\bibinfo  {journal} {Nat. Commun.}\ }\textbf {\bibinfo
  {volume} {7}},\ \bibinfo {pages} {8674} (\bibinfo {year} {2016})}\BibitemShut
  {NoStop}%
\bibitem [{\citenamefont {Sherrard}\ \emph {et~al.}(2010)\citenamefont
  {Sherrard}, \citenamefont {Robin}, \citenamefont {Lemaire},\ and\
  \citenamefont {Munro}}]{Sherrard2010}%
  \BibitemOpen
  \bibfield  {author} {\bibinfo {author} {\bibfnamefont {K.}~\bibnamefont
  {Sherrard}}, \bibinfo {author} {\bibfnamefont {F.}~\bibnamefont {Robin}},
  \bibinfo {author} {\bibfnamefont {P.}~\bibnamefont {Lemaire}}, \ and\
  \bibinfo {author} {\bibfnamefont {E.}~\bibnamefont {Munro}},\ }\href
  {\doibase 10.1016/j.cub.2010.06.075} {\bibfield  {journal} {\bibinfo
  {journal} {Curr. Biol.}\ }\textbf {\bibinfo {volume} {20}},\ \bibinfo {pages}
  {1499} (\bibinfo {year} {2010})}\BibitemShut {NoStop}%
\bibitem [{\citenamefont {Diaz de~la Loza}\ \emph {et~al.}(2018)\citenamefont
  {Diaz de~la Loza}, \citenamefont {Ray}, \citenamefont {Ganguly},
  \citenamefont {Alt}, \citenamefont {Davis}, \citenamefont {Hoppe},
  \citenamefont {Tapon}, \citenamefont {Salbreux},\ and\ \citenamefont
  {Thompson}}]{Diaz-de-la-Loza2018}%
  \BibitemOpen
  \bibfield  {author} {\bibinfo {author} {\bibfnamefont {M.-D.-C.}\
  \bibnamefont {Diaz de~la Loza}}, \bibinfo {author} {\bibfnamefont {R.~P.}\
  \bibnamefont {Ray}}, \bibinfo {author} {\bibfnamefont {P.~S.}\ \bibnamefont
  {Ganguly}}, \bibinfo {author} {\bibfnamefont {S.}~\bibnamefont {Alt}},
  \bibinfo {author} {\bibfnamefont {J.~R.}\ \bibnamefont {Davis}}, \bibinfo
  {author} {\bibfnamefont {A.}~\bibnamefont {Hoppe}}, \bibinfo {author}
  {\bibfnamefont {N.}~\bibnamefont {Tapon}}, \bibinfo {author} {\bibfnamefont
  {G.}~\bibnamefont {Salbreux}}, \ and\ \bibinfo {author} {\bibfnamefont
  {B.~J.}\ \bibnamefont {Thompson}},\ }\href {\doibase
  10.1016/j.devcel.2018.06.006} {\bibfield  {journal} {\bibinfo  {journal}
  {Dev. Cell}\ }\textbf {\bibinfo {volume} {46}},\ \bibinfo {pages} {23}
  (\bibinfo {year} {2018})}\BibitemShut {NoStop}%
\bibitem [{\citenamefont {Heller}\ \emph {et~al.}(2016)\citenamefont {Heller},
  \citenamefont {Hoppe}, \citenamefont {Restrepo}, \citenamefont {Gatti},
  \citenamefont {Tournier}, \citenamefont {Tapon}, \citenamefont {Basler},\
  and\ \citenamefont {Mao}}]{Heller2016}%
  \BibitemOpen
  \bibfield  {author} {\bibinfo {author} {\bibfnamefont {D.}~\bibnamefont
  {Heller}}, \bibinfo {author} {\bibfnamefont {A.}~\bibnamefont {Hoppe}},
  \bibinfo {author} {\bibfnamefont {S.}~\bibnamefont {Restrepo}}, \bibinfo
  {author} {\bibfnamefont {L.}~\bibnamefont {Gatti}}, \bibinfo {author}
  {\bibfnamefont {A.~L.}\ \bibnamefont {Tournier}}, \bibinfo {author}
  {\bibfnamefont {N.}~\bibnamefont {Tapon}}, \bibinfo {author} {\bibfnamefont
  {K.}~\bibnamefont {Basler}}, \ and\ \bibinfo {author} {\bibfnamefont
  {Y.}~\bibnamefont {Mao}},\ }\href {\doibase 10.1016/j.devcel.2015.12.012}
  {\bibfield  {journal} {\bibinfo  {journal} {Dev. Cell}\ }\textbf {\bibinfo
  {volume} {36}},\ \bibinfo {pages} {103} (\bibinfo {year} {2016})}\BibitemShut
  {NoStop}%
\bibitem [{\citenamefont {Ishihara}\ \emph {et~al.}(2017)\citenamefont
  {Ishihara}, \citenamefont {Marcq},\ and\ \citenamefont
  {Sugimura}}]{Ishihara2017a}%
  \BibitemOpen
  \bibfield  {author} {\bibinfo {author} {\bibfnamefont {S.}~\bibnamefont
  {Ishihara}}, \bibinfo {author} {\bibfnamefont {P.}~\bibnamefont {Marcq}}, \
  and\ \bibinfo {author} {\bibfnamefont {K.}~\bibnamefont {Sugimura}},\ }\href
  {\doibase 10.1103/PhysRevE.96.022418} {\bibfield  {journal} {\bibinfo
  {journal} {Phys. Rev. E}\ }\textbf {\bibinfo {volume} {96}},\ \bibinfo
  {pages} {022418} (\bibinfo {year} {2017})}\BibitemShut {NoStop}%
\bibitem [{\citenamefont {Latorre}\ \emph {et~al.}(2018)\citenamefont
  {Latorre}, \citenamefont {Kale}, \citenamefont {Casares}, \citenamefont
  {G{\'{o}}mez-Gonz{\'{a}}lez}, \citenamefont {Uroz}, \citenamefont {Valon},
  \citenamefont {Nair}, \citenamefont {Garreta}, \citenamefont {Montserrat},
  \citenamefont {del Campo}, \citenamefont {Ladoux}, \citenamefont {Arroyo},\
  and\ \citenamefont {Trepat}}]{Latorre2018}%
  \BibitemOpen
  \bibfield  {author} {\bibinfo {author} {\bibfnamefont {E.}~\bibnamefont
  {Latorre}}, \bibinfo {author} {\bibfnamefont {S.}~\bibnamefont {Kale}},
  \bibinfo {author} {\bibfnamefont {L.}~\bibnamefont {Casares}}, \bibinfo
  {author} {\bibfnamefont {M.}~\bibnamefont {G{\'{o}}mez-Gonz{\'{a}}lez}},
  \bibinfo {author} {\bibfnamefont {M.}~\bibnamefont {Uroz}}, \bibinfo {author}
  {\bibfnamefont {L.}~\bibnamefont {Valon}}, \bibinfo {author} {\bibfnamefont
  {R.~V.}\ \bibnamefont {Nair}}, \bibinfo {author} {\bibfnamefont
  {E.}~\bibnamefont {Garreta}}, \bibinfo {author} {\bibfnamefont
  {N.}~\bibnamefont {Montserrat}}, \bibinfo {author} {\bibfnamefont
  {A.}~\bibnamefont {del Campo}}, \bibinfo {author} {\bibfnamefont
  {B.}~\bibnamefont {Ladoux}}, \bibinfo {author} {\bibfnamefont
  {M.}~\bibnamefont {Arroyo}}, \ and\ \bibinfo {author} {\bibfnamefont
  {X.}~\bibnamefont {Trepat}},\ }\href {\doibase 10.1038/s41586-018-0671-4}
  {\bibfield  {journal} {\bibinfo  {journal} {Nature}\ }\textbf {\bibinfo
  {volume} {563}},\ \bibinfo {pages} {203} (\bibinfo {year}
  {2018})}\BibitemShut {NoStop}%
\bibitem [{\citenamefont {Graner}\ \emph {et~al.}(2008)\citenamefont {Graner},
  \citenamefont {Dollet}, \citenamefont {Raufaste},\ and\ \citenamefont
  {Marmottant}}]{Graner2008}%
  \BibitemOpen
  \bibfield  {author} {\bibinfo {author} {\bibfnamefont {F.}~\bibnamefont
  {Graner}}, \bibinfo {author} {\bibfnamefont {B.}~\bibnamefont {Dollet}},
  \bibinfo {author} {\bibfnamefont {C.}~\bibnamefont {Raufaste}}, \ and\
  \bibinfo {author} {\bibfnamefont {P.}~\bibnamefont {Marmottant}},\ }\href
  {\doibase 10.1140/epje/i2007-10298-8} {\bibfield  {journal} {\bibinfo
  {journal} {Eur. Phys. J. E}\ }\textbf {\bibinfo {volume} {25}},\ \bibinfo
  {pages} {349} (\bibinfo {year} {2008})}\BibitemShut {NoStop}%
\bibitem [{\citenamefont {Duda}\ and\ \citenamefont {Hart}(1972)}]{Duda1972}%
  \BibitemOpen
  \bibfield  {author} {\bibinfo {author} {\bibfnamefont {R.~O.}\ \bibnamefont
  {Duda}}\ and\ \bibinfo {author} {\bibfnamefont {P.~E.}\ \bibnamefont
  {Hart}},\ }\href {\doibase 10.1145/361237.361242} {\bibfield  {journal}
  {\bibinfo  {journal} {Commun. ACM}\ }\textbf {\bibinfo {volume} {15}},\
  \bibinfo {pages} {11} (\bibinfo {year} {1972})}\BibitemShut {NoStop}%
\bibitem [{\citenamefont {Streichan}\ \emph {et~al.}(2018)\citenamefont
  {Streichan}, \citenamefont {Lefebvre}, \citenamefont {Noll}, \citenamefont
  {Wieschaus},\ and\ \citenamefont {Shraiman}}]{Streichan2018a}%
  \BibitemOpen
  \bibfield  {author} {\bibinfo {author} {\bibfnamefont {S.~J.}\ \bibnamefont
  {Streichan}}, \bibinfo {author} {\bibfnamefont {M.~F.}\ \bibnamefont
  {Lefebvre}}, \bibinfo {author} {\bibfnamefont {N.}~\bibnamefont {Noll}},
  \bibinfo {author} {\bibfnamefont {E.~F.}\ \bibnamefont {Wieschaus}}, \ and\
  \bibinfo {author} {\bibfnamefont {B.~I.}\ \bibnamefont {Shraiman}},\ }\href
  {\doibase 10.7554/eLife.27454} {\bibfield  {journal} {\bibinfo  {journal}
  {Elife}\ }\textbf {\bibinfo {volume} {7}} (\bibinfo {year} {2018}),\
  10.7554/eLife.27454}\BibitemShut {NoStop}%
\bibitem [{\citenamefont {Lehoucq}\ \emph {et~al.}(2015)\citenamefont
  {Lehoucq}, \citenamefont {Weiss}, \citenamefont {Dubrulle}, \citenamefont
  {Amon}, \citenamefont {{Le Bouil}}, \citenamefont {Crassous}, \citenamefont
  {Amitrano},\ and\ \citenamefont {Graner}}]{Lehoucq2015}%
  \BibitemOpen
  \bibfield  {author} {\bibinfo {author} {\bibfnamefont {R.}~\bibnamefont
  {Lehoucq}}, \bibinfo {author} {\bibfnamefont {J.}~\bibnamefont {Weiss}},
  \bibinfo {author} {\bibfnamefont {B.}~\bibnamefont {Dubrulle}}, \bibinfo
  {author} {\bibfnamefont {A.}~\bibnamefont {Amon}}, \bibinfo {author}
  {\bibfnamefont {A.}~\bibnamefont {{Le Bouil}}}, \bibinfo {author}
  {\bibfnamefont {J.}~\bibnamefont {Crassous}}, \bibinfo {author}
  {\bibfnamefont {D.}~\bibnamefont {Amitrano}}, \ and\ \bibinfo {author}
  {\bibfnamefont {F.}~\bibnamefont {Graner}},\ }\href {\doibase
  10.3389/fphy.2014.00084} {\bibfield  {journal} {\bibinfo  {journal} {Front.
  Phys.}\ }\textbf {\bibinfo {volume} {2}},\ \bibinfo {pages} {84} (\bibinfo
  {year} {2015})}\BibitemShut {NoStop}%
\bibitem [{\citenamefont {Boudaoud}\ \emph {et~al.}(2014)\citenamefont
  {Boudaoud}, \citenamefont {Burian}, \citenamefont {Borowska-Wykret},
  \citenamefont {Uyttewaal}, \citenamefont {Wrzalik}, \citenamefont
  {Kwiatkowska},\ and\ \citenamefont {Hamant}}]{Boudaoud2014}%
  \BibitemOpen
  \bibfield  {author} {\bibinfo {author} {\bibfnamefont {A.}~\bibnamefont
  {Boudaoud}}, \bibinfo {author} {\bibfnamefont {A.}~\bibnamefont {Burian}},
  \bibinfo {author} {\bibfnamefont {D.}~\bibnamefont {Borowska-Wykret}},
  \bibinfo {author} {\bibfnamefont {M.}~\bibnamefont {Uyttewaal}}, \bibinfo
  {author} {\bibfnamefont {R.}~\bibnamefont {Wrzalik}}, \bibinfo {author}
  {\bibfnamefont {D.}~\bibnamefont {Kwiatkowska}}, \ and\ \bibinfo {author}
  {\bibfnamefont {O.}~\bibnamefont {Hamant}},\ }\href@noop {} {\bibfield
  {journal} {\bibinfo  {journal} {Nature}\ }\textbf {\bibinfo {volume} {9}},\
  \bibinfo {pages} {457} (\bibinfo {year} {2014})}\BibitemShut {NoStop}%
\bibitem [{\citenamefont {Bosveld}\ \emph {et~al.}(2012)\citenamefont
  {Bosveld}, \citenamefont {Bonnet}, \citenamefont {Guirao}, \citenamefont
  {Tlili}, \citenamefont {Wang}, \citenamefont {Petitalot}, \citenamefont
  {Marchand}, \citenamefont {Bardet}, \citenamefont {Marcq}, \citenamefont
  {Graner},\ and\ \citenamefont {Bella{\"{i}}che}}]{Bosveld2012}%
  \BibitemOpen
  \bibfield  {author} {\bibinfo {author} {\bibfnamefont {F.}~\bibnamefont
  {Bosveld}}, \bibinfo {author} {\bibfnamefont {I.}~\bibnamefont {Bonnet}},
  \bibinfo {author} {\bibfnamefont {B.}~\bibnamefont {Guirao}}, \bibinfo
  {author} {\bibfnamefont {S.}~\bibnamefont {Tlili}}, \bibinfo {author}
  {\bibfnamefont {Z.}~\bibnamefont {Wang}}, \bibinfo {author} {\bibfnamefont
  {A.}~\bibnamefont {Petitalot}}, \bibinfo {author} {\bibfnamefont
  {R.}~\bibnamefont {Marchand}}, \bibinfo {author} {\bibfnamefont {P.-L.}\
  \bibnamefont {Bardet}}, \bibinfo {author} {\bibfnamefont {P.}~\bibnamefont
  {Marcq}}, \bibinfo {author} {\bibfnamefont {F.}~\bibnamefont {Graner}}, \
  and\ \bibinfo {author} {\bibfnamefont {Y.}~\bibnamefont {Bella{\"{i}}che}},\
  }\href {http://science.sciencemag.org/content/336/6082/724.abstract}
  {\bibfield  {journal} {\bibinfo  {journal} {Science.}\ }\textbf {\bibinfo
  {volume} {336}},\ \bibinfo {pages} {724 } (\bibinfo {year}
  {2012})}\BibitemShut {NoStop}%
\bibitem [{\citenamefont {Bruns}(2004)}]{Bruns2004}%
  \BibitemOpen
  \bibfield  {author} {\bibinfo {author} {\bibfnamefont {A.}~\bibnamefont
  {Bruns}},\ }\href {\doibase 10.1016/J.JNEUMETH.2004.03.002} {\bibfield
  {journal} {\bibinfo  {journal} {J. Neurosci. Methods}\ }\textbf {\bibinfo
  {volume} {137}},\ \bibinfo {pages} {321} (\bibinfo {year}
  {2004})}\BibitemShut {NoStop}%
\bibitem [{\citenamefont {Bruns}(2005)}]{Bruns2005}%
  \BibitemOpen
  \bibfield  {author} {\bibinfo {author} {\bibfnamefont {A.}~\bibnamefont
  {Bruns}},\ }\href {\doibase 10.1016/J.JNEUMETH.2005.03.001} {\bibfield
  {journal} {\bibinfo  {journal} {Erratum, J. Neurosci. Methods}\ }\textbf
  {\bibinfo {volume} {143}},\ \bibinfo {pages} {237} (\bibinfo {year}
  {2005})}\BibitemShut {NoStop}%
\bibitem [{\citenamefont {Moisan}(2011)}]{Moisan2011}%
  \BibitemOpen
  \bibfield  {author} {\bibinfo {author} {\bibfnamefont {L.}~\bibnamefont
  {Moisan}},\ }\href {\doibase 10.1007/s10851-010-0227-1} {\bibfield  {journal}
  {\bibinfo  {journal} {J. Math. Imaging Vis.}\ }\textbf {\bibinfo {volume}
  {39}},\ \bibinfo {pages} {161} (\bibinfo {year} {2011})}\BibitemShut
  {NoStop}%
\bibitem [{\citenamefont {Bagi}(2006)}]{Bagi2006}%
  \BibitemOpen
  \bibfield  {author} {\bibinfo {author} {\bibfnamefont {K.}~\bibnamefont
  {Bagi}},\ }\href {\doibase 10.1016/J.IJSOLSTR.2005.07.016} {\bibfield
  {journal} {\bibinfo  {journal} {Int. J. Solids Struct.}\ }\textbf {\bibinfo
  {volume} {43}},\ \bibinfo {pages} {3166} (\bibinfo {year}
  {2006})}\BibitemShut {NoStop}%
\bibitem [{\citenamefont {Farahani}\ and\ \citenamefont
  {Naghdabadi}(2000)}]{Farahani2000}%
  \BibitemOpen
  \bibfield  {author} {\bibinfo {author} {\bibfnamefont {K.}~\bibnamefont
  {Farahani}}\ and\ \bibinfo {author} {\bibfnamefont {R.}~\bibnamefont
  {Naghdabadi}},\ }\href {\doibase 10.1016/S0020-7683(99)00209-7} {\bibfield
  {journal} {\bibinfo  {journal} {Int. J. Solids Struct.}\ }\textbf {\bibinfo
  {volume} {37}},\ \bibinfo {pages} {5247} (\bibinfo {year}
  {2000})}\BibitemShut {NoStop}%
\bibitem [{\citenamefont {Hencky}(1931)}]{Hencky1931}%
  \BibitemOpen
  \bibfield  {author} {\bibinfo {author} {\bibfnamefont {H.}~\bibnamefont
  {Hencky}},\ }\href {\doibase 10.1122/1.2116361} {\bibfield  {journal}
  {\bibinfo  {journal} {J. Rheol.}\ }\textbf {\bibinfo {volume} {2}},\ \bibinfo
  {pages} {169} (\bibinfo {year} {1931})}\BibitemShut {NoStop}%
\bibitem [{\citenamefont {Durande}(2019)}]{Durande:2019}%
  \BibitemOpen
  \bibfield  {author} {\bibinfo {author} {\bibfnamefont {M.}~\bibnamefont
  {Durande}},\ }\href@noop {} {\emph {\bibinfo {title} {Github codes for Fast
  determination of coarse grained cell anisotropy and size in epithelial tissue
  images using Fourier transform,
  https://github.com/mdurande/coarse-grained-anisotropy-and-size-using-FFT}}}
  (\bibinfo {year} {2019})\BibitemShut {NoStop}%
\bibitem [{\citenamefont {Hilbert}(2013)}]{Hilbert:2013}%
  \BibitemOpen
  \bibfield  {author} {\bibinfo {author} {\bibfnamefont {S.}~\bibnamefont
  {Hilbert}},\ }\href@noop {} {\emph {\bibinfo {title} {FFT Zero Padding,
  http://www.bitweenie.com/listings/fft-zero-padding/?}}} (\bibinfo {year}
  {2013})\BibitemShut {NoStop}%
\bibitem [{\citenamefont {Rozbicki}\ \emph {et~al.}(2015)\citenamefont
  {Rozbicki}, \citenamefont {Chuai}, \citenamefont {Karjalainen}, \citenamefont
  {Song}, \citenamefont {Sang}, \citenamefont {Martin}, \citenamefont
  {Kn{\"{o}}lker}, \citenamefont {MacDonald},\ and\ \citenamefont
  {Weijer}}]{Rozbicki2015}%
  \BibitemOpen
  \bibfield  {author} {\bibinfo {author} {\bibfnamefont {E.}~\bibnamefont
  {Rozbicki}}, \bibinfo {author} {\bibfnamefont {M.}~\bibnamefont {Chuai}},
  \bibinfo {author} {\bibfnamefont {A.~I.}\ \bibnamefont {Karjalainen}},
  \bibinfo {author} {\bibfnamefont {F.}~\bibnamefont {Song}}, \bibinfo {author}
  {\bibfnamefont {H.~M.}\ \bibnamefont {Sang}}, \bibinfo {author}
  {\bibfnamefont {R.}~\bibnamefont {Martin}}, \bibinfo {author} {\bibfnamefont
  {H.-J.}\ \bibnamefont {Kn{\"{o}}lker}}, \bibinfo {author} {\bibfnamefont
  {M.~P.}\ \bibnamefont {MacDonald}}, \ and\ \bibinfo {author} {\bibfnamefont
  {C.~J.}\ \bibnamefont {Weijer}},\ }\href {\doibase 10.1038/ncb3138}
  {\bibfield  {journal} {\bibinfo  {journal} {Nat. Cell Biol.}\ }\textbf
  {\bibinfo {volume} {17}},\ \bibinfo {pages} {397} (\bibinfo {year}
  {2015})}\BibitemShut {NoStop}%
\bibitem [{\citenamefont {Tlili}\ \emph {et~al.}(2018)\citenamefont {Tlili},
  \citenamefont {Gay}, \citenamefont {Ladoux}, \citenamefont {Graner},\ and\
  \citenamefont {Delano{\"{e}}-Ayari}}]{Tlili2018b}%
  \BibitemOpen
  \bibfield  {author} {\bibinfo {author} {\bibfnamefont {S.}~\bibnamefont
  {Tlili}}, \bibinfo {author} {\bibfnamefont {C.}~\bibnamefont {Gay}}, \bibinfo
  {author} {\bibfnamefont {B.}~\bibnamefont {Ladoux}}, \bibinfo {author}
  {\bibfnamefont {F.}~\bibnamefont {Graner}}, \ and\ \bibinfo {author}
  {\bibfnamefont {H.}~\bibnamefont {Delano{\"{e}}-Ayari}},\ }\href
  {http://arxiv.org/abs/1811.05001} {\  (\bibinfo {year} {2018})},\ \Eprint
  {http://arxiv.org/abs/1811.05001} {arXiv:1811.05001} \BibitemShut {NoStop}%
\bibitem [{\citenamefont {Tanner}\ and\ \citenamefont
  {Tanner}(2003)}]{Tanner2003}%
  \BibitemOpen
  \bibfield  {author} {\bibinfo {author} {\bibfnamefont {R.}~\bibnamefont
  {Tanner}}\ and\ \bibinfo {author} {\bibfnamefont {E.}~\bibnamefont
  {Tanner}},\ }\href {\doibase 10.1007/s00397-002-0259-6} {\bibfield  {journal}
  {\bibinfo  {journal} {Rheol. Acta}\ }\textbf {\bibinfo {volume} {42}},\
  \bibinfo {pages} {93} (\bibinfo {year} {2003})}\BibitemShut {NoStop}%
\end{thebibliography}%

\clearpage
\begin{appendix}
\section{Matrices used in the text}
\label{sec:def}

We introduce here three types of 2$\times$2 matrices, also called rank-2 tensors: the inertia matrix, the FT inertia matrix and the cell strain.

The inertia matrix of a binarized pattern is defined by: 
\begin{equation}
\label{eq:inertia}
I = \begin{pmatrix}
\left\langle xx \right\rangle &  \left\langle xy  \right\rangle  \\
\left\langle  xy  \right\rangle  & \left\langle  yy \right\rangle  
\end{pmatrix}
\end{equation}
Here $x$ and $y$ are coordinates with origin at the pattern barycenter, $\left\langle x \right\rangle =  \left\langle y  \right\rangle = 0$.
The brackets indicate an average over the coordinates within the pattern 
(if the pattern was in grey levels instead of being binarized, the average would be weighted by the grey levels). 
The four terms which appear in $I$ are the coordinate covariances.

The Fourier transform inertia matrix has the same definition, Eq.~(\ref{eq:inertia}), but it operates in the Fourier space. Here $x$ and $y$ are coordinates in the space of spatial frequencies, again with origin at the pattern barycenter, $\left\langle x \right\rangle =  \left\langle y  \right\rangle = 0$.


\begin{figure}[h!]
\includegraphics[scale = 1]{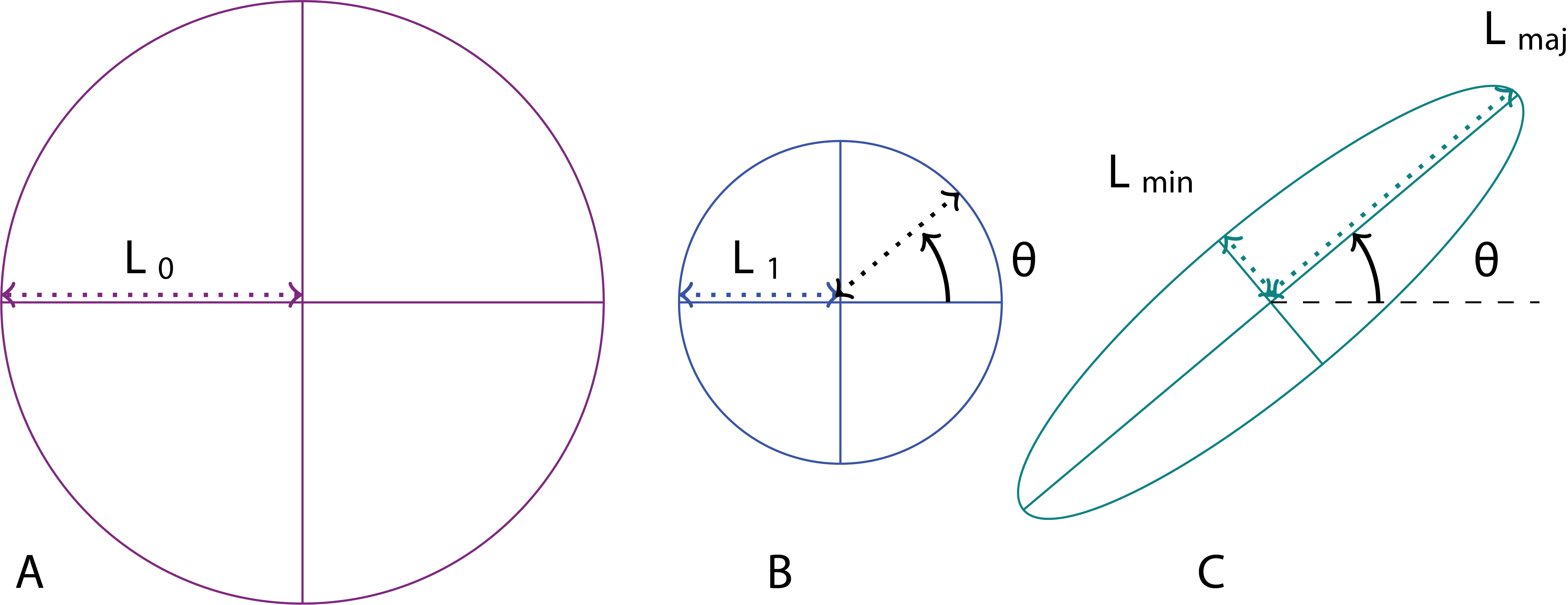}
\caption{Strain: isotropic and anisotropic contributions. 
Under a purely isotropic deformation, or growth (positive or negative), a disk of radius $L_0$ $(A)$ transforms into a disk of radius $L_1$ $(B)$.
Under a purely anisotropic deformation, or convergence-extension, a disk of radius $L_1$ $(B)$ transforms into an ellipse of major axis $L_{maj}$, in direction $\theta$, and minor axis $L_{min}$, in perpendicular direction, with area conservation expressed by  $L_{maj}L_{min}=L_1^2$ $(C)$.
}
\label{im:deftic}
\end{figure}

The cell strain has isotropic and anisotropic contributions  (Fig.~\ref{im:deftic})
\begin{equation}
\varepsilon_c = \frac{1}{2}{\mathrm{Tr}}(\varepsilon_c)\mathcal{I}_2 + \varepsilon_c^{dev}
\end{equation}
where $\mathcal{I}_2$ is the identity matrix in 2 dimensions \modif{and $\mathrm{Tr}$ is the trace}.

\begin{figure}[h!]
\includegraphics[scale=1]{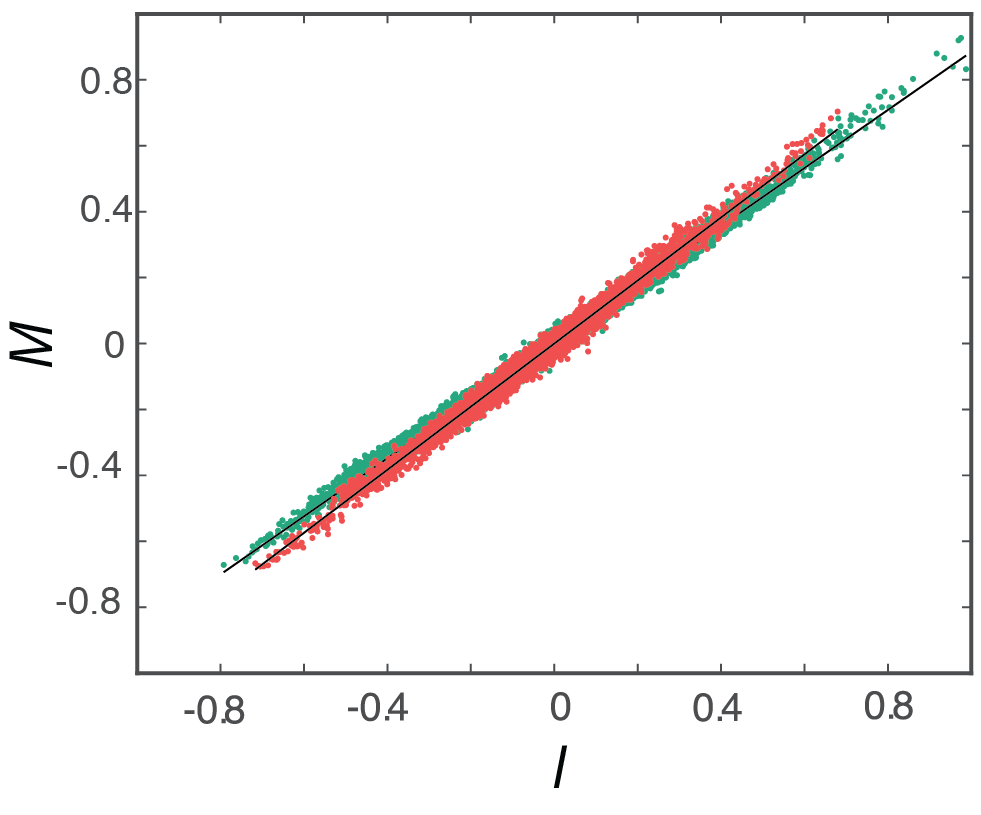}
\caption{Texture \modif{tensor} versus inertia matrix. Data presented here are from the same dataset as Fig. \ref{fig:pup}. Diagonal component (green) and off-diagonal component (red) of the texture tensor $M$ as defined in \modif{Ref.}\cite{Graner2008} versus the corresponding anisotropic (diagonal and off-diagonal) component of the inertia matrix $I$, representing cell shape anisotropy. Slopes are 0.88 \modif{and} 0.96, correlations coefficients are  $0.99$ and $0.98$, respectively. Tensors built with data from  \cite{Guirao2015}, here plotted after adimensionalisation by the isotropic part of the respective tensor. To compute each of the 14112 points, tensors of individual cells are computed \modif{then} averaged in Eulerian grids of $40 \times 40$~$\mu$m$^2$ with 50\% overlap. Then a sliding average is performed on 2 h (24 frames) of time with a one hour overlap. Boxes at the pattern boundary which are filled at less than 30\% by cells are excluded from the fit.}
\label{fig:MI}
\end{figure}

Consider a circle of radius $\ell$ (Fig.~\ref{im:deftic}) and apply a small variation of its length $d\ell$. Its relative extension is $d\ell/\ell = d(\log{\ell})$. Integrating this infinitesimal extension between the initial and final states yields the expression for strain \cite{Tanner2003} which for the isotropic part writes (Fig.~\ref{im:deftic}A,B):
\begin{equation}
\frac{1}{2}{\mathrm{Tr}}(\varepsilon_c)\mathcal{I}_2 
=
 \frac{\log \left(L_1 / L_0 \right)}{2}
\begin{pmatrix}
1 & 0    \\
0  &  1  
\end{pmatrix} 
\end{equation}
and for the anisotropic part, after diagonalisation along axes of directions $\theta$ and $\theta + \pi/2$ (Fig.~\ref{im:deftic}B,C):
\begin{equation}
\varepsilon_c^{dev}
=
\begin{pmatrix}
\log \left(\frac{ L_{maj}}{L_1}\right)& 0    \\
0  &  \log \left(\frac{ L_{min}}{L_1}\right) 
\end{pmatrix} 
=
\begin{pmatrix}
\frac{ \log \left(L_{maj}/L_{min}\right)}{2} & 0    \\
0  &  - \frac{ \log \left(L_{maj}/L_{min}\right)}{2}   
\end{pmatrix} 
\end{equation}
where $L_1 = \sqrt{L_{maj}L_{min}}$.
The cell strain deviator amplitude is $ \frac{1}{2}\log{\frac{L_{maj}}{L_{min}}} $, and the cell strain deviator orientation is $\theta$. The cell strain deviator can be inferred from the pattern anisotropy, without requiring any information about the current cell size $L_1$ or its rest state value $L_0$; the above derivation only assumes that the rest state is isotropic. Note that this definition of the strain is called the ``true" strain, or Hencky strain \cite{Hencky1931}.
When the cell strain deviator amplitude is much smaller than one, one can alternatively use any approximation equivalent at first order, such as $ \frac{1}{2}\left(\frac{L_{maj}}{L_{min}} - 1 \right)$, for instance when using the linear approximation to the true strain,
or $ \frac{1}{4}\left(\frac{L^2_{maj}}{L^2_{min}} - 1 \right)$, when using quadratic tensors attached to the matter\modif{: inertia matrix, defined in Eq.~(\ref{eq:inertia}), or texture tensor, defined in Ref. \cite{Graner2008}.} \modif{We have also checked that the inertia matrix and the texture tensor statistically yield equivalent information (Fig. \ref{fig:MI}).}

\newpage

\section{Robustness of the inertia matrix method }

\begin{figure}[h!]
\includegraphics[scale=0.8]{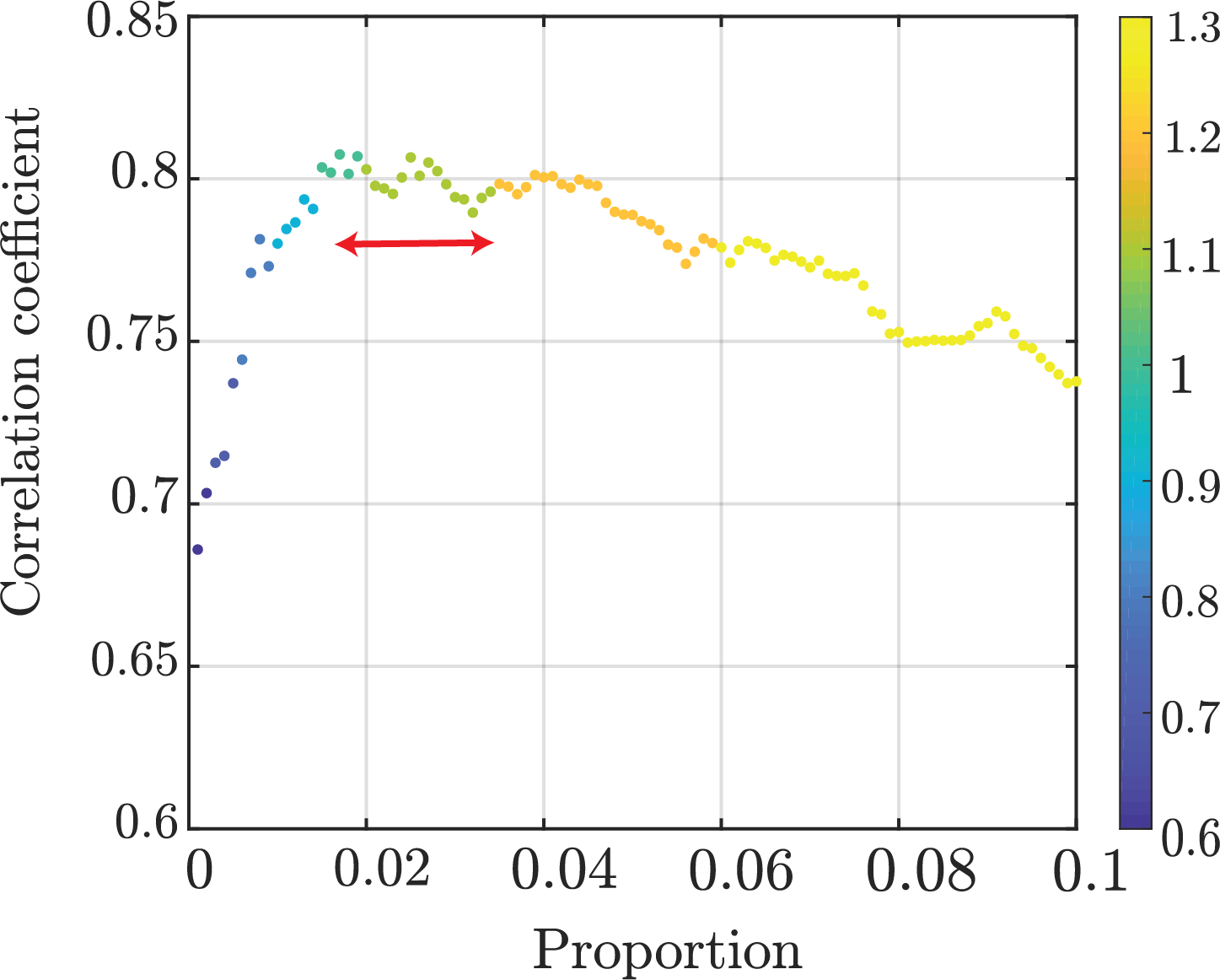}
\caption{\modif{Robustness of  the inertia matrix method quality versus choice of parameter: correlation coefficient  versus ``proportion" parameter (see text), for the Drosophila pupa data set. The correlation slope is color coded. The method quality is optimal when the correlation coefficient is high and the slope close to 1. The red arrow represents the parameter range which can reasonably be used for this data set.}}
\label{im:corrcoef}
\end{figure}

\modif{Fig. \ref{im:corrcoef} investigates the robustness of  the inertia matrix method quality versus the choice of the main parameter, the manually selected \modif{proportion $p$} of bright pixels used when thresholding the Fourier transform (Fig.~\ref{im:imMethod}H).
Using the Drosophila pupa dataset, for each value of the proportion  we compare the inertia matrix method  results with the segmentation analysis considered as a gold standard, by performing a linear regression on data with anisotropy larger than 0.08. 
The method quality is considered as optimal when the linear regression has a slope close to 1 and its correlation coefficient is high.  
We find an optimum for a proportion around $2 \cdot 10^{-2}$
and a large parameter range around this value where the method quality is robust (red arrow).}

\label{sec:paramrange}

\end{appendix}

\end{document}